\documentclass[usenatbib]{mn2e}

\usepackage{psfig,morefloats,url}

\footnotesize
\newdimen\digitwidth    
\setbox0=\hbox{\rm0}
\digitwidth=\wd0
\catcode`!=\active
\def!{\kern\digitwidth}
\normalsize

\title[The Parkes multibeam pulsar survey: VI.] {The Parkes multibeam
pulsar survey: VI. Discovery and timing of 142 pulsars and a
Galactic population analysis}
\author[D.~R.~Lorimer et al.]
{D.~R.~Lorimer$^{1,2}$,\thanks{Email: Duncan.Lorimer@mail.wvu.edu}
A.~J.~Faulkner$^1$, A.~G.~Lyne$^1$, R.~N.~Manchester$^3$, M.~Kramer$^1$,
\newauthor
M.~A. McLaughlin$^{1,2}$, G.~Hobbs$^3$, A.~Possenti$^4$, I.~H.~Stairs$^5$, 
F.~Camilo$^6$,
\newauthor
M.~Burgay$^4$, N.~D'Amico$^{4,7}$, A.~Corongiu$^{4}$ and F.~Crawford$^8$
\\
$^1$ University of Manchester, Jodrell Bank Observatory, Macclesfield,
Cheshire SK11~9DL \\
$^2$ Department of Physics, West Virginia University, PO~Box~6315, Morgantown,
WV~26506, USA\\
$^3$ Australia Telescope National Facility, CSIRO, PO~Box~76, Epping
NSW~1710, Australia \\
$^4$ INAF - Osservatorio Astronomico di Cagliari, Loc. Poggio dei Pini,
Strada 54, 09012, Capoterra (CA), Italy \\
$^5$ Department of Physics \& Astronomy, University of British Columbia,
6224 Agricultural Road, Vancouver, B.C. V6T 1Z1, Canada \\
$^6$ Columbia Astrophysics Laboratory, Columbia University, 550 West 
120th Street, New York, NY 10027, USA \\
$^7$ Universita' degli Studi di Cagliari, Dipartimento di Fisica, SP
   Monserrato-Sestu km 0,7, 90042, Monserrato (CA), Italy \\
$^8$ Department of Physics and Astronomy, Franklin \& Marshall College,
PO~Box~3003, Lancaster, PA 17604, USA}
\let\leqslant=\leq 
\date{Accepted for publication in MNRAS}
\begin{document}

\maketitle
\newcommand{\setthebls}{
}

\setthebls

\begin{abstract} 
We present the discovery and follow-up observations of 142 pulsars
found in the Parkes 20-cm multibeam pulsar survey of the Galactic plane.
These new discoveries bring the total number of
pulsars found by the survey to 742.  In addition to tabulating spin
and astrometric parameters, along with pulse width and flux density
information, we present orbital characteristics for 13 binary pulsars
which form part of the new sample. Combining these results from
another recent Parkes multibeam survey at high Galactic latitudes, we
have a sample of 1008 normal pulsars which we use to carry out a
determination of their Galactic distribution and birth rate. We infer
a total Galactic population of $30000\pm 1100$ potentially detectable
pulsars (i.e.~those beaming towards us) having 1.4-GHz luminosities
above 0.1~mJy~kpc$^2$. Adopting the Tauris \& Manchester beaming
model, this translates to a total of $155000\pm6000$ active radio
pulsars in the Galaxy above this luminosity limit. Using a pulsar
current analysis, we derive the birth rate of this population to be
$1.4 \pm 0.2$ pulsars per century. An important conclusion from our
work is that the inferred radial density function of pulsars depends
strongly on the assumed distribution of free electrons in the
Galaxy. As a result, any analyses using the most recent electron model
of Cordes \& Lazio predict a dearth of pulsars in the inner Galaxy. We
show that this model can also bias the inferred pulsar scale height
with respect to the Galactic plane. Combining our results with other
Parkes multibeam surveys we find that the population is best described
by an exponential distribution with a scale height of 330 pc. Surveys
underway at Parkes and Arecibo are expected to improve the
knowledge of the radial distribution outside the solar circle, and
to discover several hundred new pulsars in the inner Galaxy.
\end{abstract}


\section{INTRODUCTION}

The Parkes multibeam pulsar survey of the Galactic plane is the most
successful large scale search for pulsars so far undertaken.  Five
previous papers in this series have presented timing parameters for
600 newly discovered pulsars and have discussed various aspects of the
survey results. A detailed description of the 13-beam 20-cm receiver system
and data reduction software can be found in \citet{mlc+01}.
Preliminary remarks about the population statistics were made by
\citet{mhl+02}.  \citet{kbm+03} discussed the association of young
pulsars with unidentified EGRET sources and gave evidence for pulsars
tracing the spiral-arm structure of the Galaxy. \citet{hfs+04}
discussed the detection of previously known pulsars in the
survey. Finally, \citet{fsk+04} implemented and discussed improved
processing and candidate selection strategies for the entire survey
database with an emphasis on finding binary and millisecond pulsars.

In this paper, we present the discovery parameters and provide timing
solutions for a further 142 new pulsars.  This brings the total number
of pulsars found in the survey to 742.  Although future searches of
the data may produce a few additional discoveries, the vast majority
of ``normal'' (i.e.~non-recycled) pulsars detectable by the survey
have now been found.  It is therefore appropriate to use this sample
to place new constraints on the Galactic distribution and birth rate
of normal pulsars. Of particular interest is the distribution in
Galactocentric radius. Due to severe selection effects on
low-frequency ($<1$~GHz) pulsar surveys, earlier studies \citep[see,
e.g.,][]{tm77} were somewhat hampered by small number statistics in
the inner Galaxy and were often limited to local population analyses
\citep[see, e.g.,][]{lml+98}. In recent years, with the discovery of a
larger number of pulsars in 1.4-GHz surveys there has been a growing
body of evidence which suggests a deficit of pulsars in the inner
Galaxy relative to a simple model where the radial density profile
follows a Gaussian distribution \citep{joh90,yk04,lor04}. One of the
main goals of this paper is to use the new sample, which provides
greatly improved pulsar statistics in the inner Galaxy, to provide an
updated analysis on the Galactic distribution.

The plan for this paper is as follows.  In \S\ref{sec:timing} we
present the basic timing parameters, pulse widths, mean profiles and
flux densities for the 142 new pulsars. The sample contains a number
of interesting individual objects, including several binary and
millisecond pulsars which are discussed in \S\ref{sec:individual}.  In
\S\ref{sec:psrpop} we describe our Galactic population analysis. The
main conclusions from this work are summarized in \S\ref{sec:conc}.

\section{DISCOVERY AND TIMING OF 142 PULSARS}\label{sec:timing}

The pulsars presented here were discovered primarily using the
processing schemes described by \citet{fsk+04}. In brief, to
compensate for the dispersive effects of the interstellar medium, data
from each telescope beam were de-dispersed at 325 different trial
values of dispersion measure (DM) spanning the range 0 to 2203
cm$^{-3}$~pc. The de-dispersed time series were then subjected in turn
to: a standard fast Fourier transform (FFT) analysis to search for
periodic signals \citep{mlc+01}, a segmented FFT search for
accelerated signals from binary pulsars \citep{fsk+04}, a fast-folding
algorithm (Kramer et al.,~in preparation) primarily sensitive to
periods in the range 1--10~s and a search for dispersed single pulses
\citep{mll+06}.  Pulsar candidates from all but the latter search were
selected using a graphical tool which displayed various aspects of the
search parameter space, e.g.~signal-to-noise ratio (S/N) versus pulse
period. Promising candidates were noted for later re-observation at
Parkes.

Following the confirmation and positional refinement procedures
described by \citet{mhl+02}, each pulsar was observed regularly for at
least one year using one or more of the Parkes, Lovell and Arecibo
telescopes.  The observing systems used at Parkes and at Jodrell Bank
are described by \citet{mlc+01} and \citet{mhl+02} respectively.  The
Arecibo timing is described by \citet{hfs+04} and \citet{sfl+05}.  For
each pulsar, pulse times of arrival (TOAs) were determined from the
individual observations using standard pulsar timing techniques
\citep[see, e.g.,][]{lk05} implemented in the \textsc{psrchive}
software package
\citep{hvm04}\footnote{\url{http://psrchive.sourceforge.net}}. A model
containing the spin, astrometric and (if necessary) any binary
parameters was fitted to the TOAs using the \textsc{tempo} timing
package\footnote{\url{http://www.atnf.csiro.au/research/pulsar/tempo}}.

The positional information from these fits is provided in equatorial
and Galactic coordinates in Table~\ref{tb:posn}. Subsequent columns in
this table contain information on the discovery of each pulsar: the
beam number (corresponding to the 13 beams of the multibeam receiver)
for the strongest discovery observation, the radial distance between
the centre of this beam and the position of the pulsar (distances
greater than one beam radius can occur if the pulsar scintillates or
nulls or the closest pointing was contaminated by interference) and
the S/N of the profile\footnote{Throughout
this paper we discuss S/N measurements made in the time domain,
i.e.~determined directly from the integrated pulse profile \citep[for
further details, see][]{lk05}} during this observation.  The
observations used to form TOAs were added together to provide a
characteristic pulse profile for each pulsar at 1400\,MHz
(Figure~\ref{fg:prf}). The final three columns in Table~\ref{tb:posn}
contain the flux densities measured from these mean profiles
\citep[for further details, see][]{mlc+01}, pulse widths at 50\% and
10\% of the pulse height. The 10\% width is not measurable for pulsars
with mean profiles that have poor S/N.  For profiles containing
multiple components the widths are measured across the entire profile.

The rotational parameters from the timing analyses are given in
Table~\ref{tb:prd}.  In column order, this table provides each
pulsar's name, Solar-system barycentric pulse period, period
derivative, epoch of pulse period, the number of TOAs used in the
timing solution, the MJD range covered by the timing observations,
root-mean-square value of the observed$-$model residuals and the DM.
The data have been folded in turn at two and three times the tabulated
periods to confirm that they represent the fundamental periods of the
pulsars, rather than a harmonic.

Various derived parameters for the new pulsars are presented in
Table~\ref{tb:deriv}. For each pulsar, we list the base-10 logarithms
of the characteristic age, surface dipole magnetic field strength and
rate of loss of rotational energy.  The final columns contain the
pulsar distances, height above the Galactic plane and
luminosities\footnote{Note that throughout this paper we will use the
term luminosity $L$ to describe the quantity $L=Sd^2$, where $S$ is
the flux density and $d$ is the distance. As this omits any
geometrical and beaming factors, this quantity is often referred to as
a pseudoluminosity \citep[see, e.g.,][]{acc02}.}.  The distances are
computed from their DMs assuming the \citet{tc93} and \citet{cl02}
models for the Galactic distribution of free electrons. The former is
used for consistency with earlier papers in this series. The latter is
more up-to-date and used for the population analysis in
\S\ref{sec:gdist}.

\section{Discussion of individual objects}\label{sec:individual}

The 7.7-s pulsar J1001$-$5939 has the longest spin period in our
sample, and is the second longest known for a radio pulsar after the
8.5-s pulsar J2144$-$3933 \citep{ymj99}. Unlike PSR~J2144$-$3933, the
much larger period derivative measured for J1001$-$5939 means that the
pulsar lies above the ``death line'' in the $P-\dot{P}$ diagram
\citep{rs75}. However, its implied dipole surface magnetic field is
below the quantum critical limit \citep{bh01}.  As a result,
PSR~J1001$-$5939 does not pose any serious problems to theories of
radio emission.  Although the initial survey detection of this pulsar
was registered in its second harmonic in the standard FFT search, its
detection was much more significant using the fast-folding
algorithm. Full details of this aspect of the survey data processing
will appear elsewhere (Kramer et al.,~in preparation).

In addition to the new discoveries of rotating radio transient sources
in the single-pulse search described by \citet{mll+06}, one further
pulsar, PSR J1624$-$4613, was only detectable in this part of the
analysis. Our follow-up observations show that this pulsar is only
detectable about 70\% of the time. This and the lack of detection by
the periodicity analysis suggests that this pulsar spends a
significant fraction of time in a null state. The discovery of such
pulsars alongside the radio transients highlights the value of the
single-pulse search analysis as an important component of pulsar
survey pipelines.

The youngest pulsar in our sample, PSR~J1357$-$6429, has been
extensively discussed by \citet{cml+04}. This pulsar, which has
suffered a Vela-like glitch, is possibly associated with the radio
supernova remnant G309.8$-$2.6. We have searched for coincidences
between the newly discovered pulsars that lie within the bounds of the
231 supernova remnants in Green's catalogue as well as the 35 remnants
recently found in the inner Galaxy \citep{bgg+06}. Although five
additional matches were found (G4.2$-$3.5 and J1808$-$2701; G27.8+0.6
and J1839$-$0436; G321.9$-$0.3 and J1519$-$5734; G343.0$-$6.0 and
J1718$-$4539; and G343.0$-$6.0 J1724$-$4500) none of the pulsars have
characteristic ages below $6\times10^5$~yr or an angular offset less
than 0.45 remnant radii. Such apparent coincidences are common and
expected, given the high density of pulsars and supernova remnants
along the Galactic plane \citep{gj95b,llc98}. To demonstrate this, we
artificially shifted the remnant centres by $1\degr$ in Galactic
longitude. The subsequent fake cross-correlation analysis resulted in
6 matches with similarly old pulsars and large angular offsets. 
We therefore do not believe any of the above coincidences are
genuine pulsar--supernova remnant associations.

Several of the profiles shown in Fig.~\ref{fg:prf} exhibit the
characteristic hallmark of interstellar scattering, i.e.~a one-sided
exponential tail. The most extreme example in the current sample is
PSR~J1901+0435, a 691-ms pulsar located at Galactic longitude
$l=38^{\circ}$, which shows a scattering tail encompassing most of its
rotational phase. Further study of this and other pulsars in the
sample \citep[see, e.g.,][]{bcc+04} will help improve models of the
distribution of scattering material in the interstellar medium. These
detections of highly scattered pulsars also suggest that, even at the
survey frequency of 1.4~GHz, a number of nominally bright pulsars are
not detectable due to scattering. Further surveys of the Galactic
plane at even higher observing frequencies are required to detect
these pulsars.

Our sample includes the two solitary millisecond pulsars J1801$-$1417
and J1911$+$1347, both originally announced by \citet{fsk+04}.
Improved timing parameters for these pulsars are given in
Tables~\ref{tb:posn} and \ref{tb:prd}. These discoveries bring the
sample of solitary millisecond pulsars known in the Galactic disk to
15. In keeping with the other isolated millisecond pulsars, these new
sources have low radio luminosities compared to their counterparts
with binary companions \citep{bjb+97,kxl+98}. As noted recently by
\citet{lkn+06} the difference in scale heights of the binary and
isolated pulsars, despite their having statistically identical
velocity dispersions, also points to a difference in luminosities for
the two populations.  It is currently unclear why the luminosity
functions differ for the two populations.

Our sample includes 13 binary pulsars, the properties of which are
summarized in Table~\ref{tb:allbin}. Of these, seven are discussed
elsewhere: PSR~J1638$-$4725 is a long-period binary pulsar in a highly
eccentric orbit around a massive companion (Lyne et al., in
preparation), PSR~J1744$-$3922 is a short-period binary system with a
low-mass companion \citep{fsk+04}, PSR J1802$-$2124 has a companion of
intermediate mass \citep{fsk+04}, PSR J1756$-$2251 is a
double-neutron-star binary \citep{fkl+05} and PSRs J1751$-$2857,
J1853+1303 and J1910+1256 are millisecond pulsars in wide binary
orbits \citep{sfl+05}. In Tables~\ref{tb:btbin} and \ref{tb:elbin} we
provide orbital parameters for the remaining six binary systems.

PSRs J1711$-$4322 and J1822$-$0848 listed in Table~\ref{tb:btbin}
appear to be similar in character to PSR~B0820+02; i.e.~a long-period
pulsar in a long-period orbit about a low-mass white dwarf companion.
PSRs J1125$-$6014, J1216$-$6410 and J1841+0130 are typical of other
recycled pulsar binary systems with orbital periods of typically a few
days and low-mass white dwarf companions.  All of these binaries
follow the spin period versus orbital period and magnetic field versus
orbital period correlations discussed by \citet{vb95} which support
the largely empirical hypothesis that accretion onto the neutron star
causes a reduction in the magnetic field strength \citep[see,
e.g.,][]{bv91}.

Another important tracer for the evolution of binary pulsars is the
orbital period versus eccentricity correlation. According to the
fluctuation-dissipation theory of \citet{phi92}, a relationship
between these parameters is expected for all pulsar--white dwarf
binary systems which have undergone an extensive period of mass
transfer. With the exception of PSRs J1841+0130 and J1853+1303, which
deviate from the theoretical prediction by over an order of magnitude,
the pulsar--white dwarf systems in our sample follow the $P_b-e$
correlation. As noted by \citet{sfl+05} these two exceptions may
simply be indicative of a natural scatter about the relationship.
Further work is required to establish whether these deviations are
indicative of a different evolutionary channel, such as the formation
of the neutron star via accretion induced collapse of a white dwarf in
an accreting binary system \citep[see, e.g.,][]{np89}.

Slightly less typical is the 28-ms pulsar, PSR J1439$-$5501. This
binary pulsar with a circular orbit of period 2~days has a minimum
companion mass of 1.1~$M_{\odot}$. It appears to be among the growing
class of recycled pulsars with more massive CO white dwarf companions,
the so-called ``intermediate-mass binary pulsars'' \citep{cnst96}.
Its closest known counterpart among the known systems is the 43-ms
binary pulsar J1157$-$5112 \citep{eb01}. Another likely
intermediate-mass system is J1802$-$2124 for which the minimum
companion mass is 0.8~$M_{\odot}$ \citep{fsk+04}.

Related to the binary pulsars is the isolated pulsar PSR~J1753$-$1914.
With a spin period of 63~ms and a spin-down rate of only
$2\times10^{-18}$, this pulsar appears to belong to the class of
so-called ``disrupted recycled pulsars'' \citep{lma+04}. Since these
objects have spin properties similar to the recycled pulsars in 
double-neutron-star binary systems, it was proposed by \citet{cnt93} that
they are the result of the systems which disrupted at the time of the
second supernova. As discussed by \citet{lma+04}, there appears to be
a deficit of these systems among the known pulsar sample.  Further
investigations, especially using binary population syntheses, are
required to investigate the frequency of these isolated recycled
pulsars relative to double neutron star systems.

\section{The Galactic pulsar population}\label{sec:psrpop}

To date, the total number of pulsars discovered by the Parkes
multibeam (hereafter PMB) survey is 742.  This substantial haul,
together with a further 11 transient radio sources found in the survey
data associated with rotating neutron stars \citep{mll+06}, means that
our survey is a phenomenal new probe of the Galactic neutron-star
population. We also make use of the recently published high-latitude
pulsar survey of the region $-140^{\circ}<l<-100^{\circ}$ and
$|b|<60^{\circ}$ \citep[hereafter PH survey,][]{bjd+06}.  Since the
PMB and PH surveys were carried out using the same observing system,
the pulsars detected represent a reasonably homogeneous sample from
which to proceed.

Although the total number of pulsars detected by the PMB and PH
surveys is 1055, for this analysis, where we are concerned with the
properties of normal pulsars, we use only those 1008 non-binary
pulsars with $P>30$~ms. In this sample, 976 pulsars were detected
solely by the PMB survey, 22 were detected solely by the PH survey and
12 were detected jointly.  While there are two additional large-scale
surveys using the Parkes multibeam system \citep{ebvb01,jac05}, these
are not included at this stage of the analysis due to lack of flux
densities for many of the pulsars from these surveys. However, these
surveys are utilized later in the discussion (\S \ref{sec:z}) when the
scale height of the population is investigated.

In the following, we investigate various aspects of the Galactic
pulsar population after accounting, as far as possible, for the known
selection effects.  We begin by presenting a model for the survey
detection process which is in good agreement with the observed S/Ns
from the PMB survey.  We then use this detection model in a Monte
Carlo simulation to iteratively deduce various underlying Galactic
population properties.  Finally, we apply a pulsar current analysis on
our optimal model to deduce the birth rate of the population.

\subsection{Survey detection model}\label{sec:sdmodel}

Although \citet{mlc+01} presented a detailed formalism to calculate
the survey sensitivity, we have found that a simpler approach is to
use the radiometer equation \citep[see, e.g.,][]{dss+84}. As we show
below this provides a self-consistent model for the detection
statistics. This approach was also adopted by \citet{fk06} where it
was noted that there appears to be some inconsistencies with the
procedure given by \citet{mlc+01}. The radiometer equation gives the
observed S/N in terms of various pulsar and system parameters as follows:
\begin{equation}
\label{equ:dewey}
  {\rm S/N} = S_{1400} G \frac{\sqrt{n_p \Delta \nu \tau}}{\beta T}
  \sqrt{\frac{P-W}{W}}.
\end{equation}
Here $S_{1400}$ is the pulsar mean flux density at 1400~MHz (mJy), $G$
is the effective telescope gain (K/Jy), $n_p$ is the number of
polarizations summed, $\Delta \nu$ is the observing bandwidth (MHz),
$\tau$ is the integration time (s), $\beta$ accounts for S/N losses,
$T$ is the system temperature (K), $P$ is the pulse period (s) and $W$
is the observed pulse width (s).  Since both the PMB and PH surveys
used the same telescope, receiver and data acquisition system, the
only parameter that is different is the integration time $\tau=2097$~s
for the PMB survey and 262~s for the PH survey.  The constant
parameters are $n_p=2$, $\Delta \nu = 288$~MHz and $\beta =
\sqrt{\pi/2}\simeq1.25$ for S/N losses due to one bit digitization
\citep[see, e.g.,][]{lk05}.  Gain and system temperature values depend
primarily on which element of the 13-beam receiver system was used to
make the detection. We use the values given by \citet{mlc+01}. Sky
temperature values vary as a function of Galactic latitude and
longitude. To account for this, we use the \citet{hssw82} 430-MHz
all-sky catalogue scaled to the observing frequency of 1400 MHz
assuming a sky background spectral index of $-$2.6 \citep{lmop87}.

To account for the drop in flux density away from the beam centre, we
can assume a Gaussian beam \citep[see, e.g.,][]{lbdh93} to write
\begin{equation}
\label{equ:fluxcor}
  G = G_0 \exp(-2.77 r^2 / w^2),
\end{equation}
where $G_0$ is the telescope gain at the beam centre, $r$ is the
offset from the beam centre and $w$ is the full-width at half power of
the telescope beam. For each detection under consideration, we adopt
the $G_0$ and $w$ values given by \citet{mlc+01} for the appropriate
beam of the multibeam receiver when making this correction.  To
calculate the equivalent pulse width $W$ from the observed 50\% width,
$W_{50}$, we make a simplification that the pulse shape is Gaussian in
form (a reasonable approximation for most pulsars). For a Gaussian
pulse with an intensity of unity and a standard deviation $\sigma$, we
find that $W_{50}=\sigma \sqrt{8 \ln 2}$ and $W$ is just the area
under the pulse (i.e. $W=\sigma \sqrt{2 \pi}$). Eliminating $\sigma$
from these two equalities yields the relationship
\begin{equation}
  W = W_{50} \sqrt{\frac{\pi}{4\ln2}} \simeq 1.06 W_{50}.
\end{equation}
Figure~\ref{fg:modelsn} shows a comparison between the modeled and
observed S/N values using data taken from Table~\ref{tb:posn} and the
corresponding data from the earlier papers in this series. The error
bars are derived directly from the uncertainties in the measured flux
densities of each pulsar. As can be seen, the agreement between our
model and observed S/N values is good, being largely free from
systematic trends and with a scatter that is not much larger than the
uncertainties. In addition to providing a sanity check on the system
performance, this analysis demonstrates that Equation \ref{equ:dewey}
provides a good model of the survey detection thresholds and, for the
sample of normal pulsars presented here, we recommend its use in
future analyses.

\subsection{Galactic distribution determination method}\label{sec:gdist}

Using the above model for the PMB and PH surveys, we now describe a
method to derive the underlying probability density functions (PDFs)
in pulse period ($P$), 1400-MHz radio luminosity ($L$), Galactocentric
radius ($R$) and height above the Galactic plane ($z$). The modeling
technique we use is based on the iterative procedure developed by
\citet{lar71}, \citet{tm77} and \citet{lmt85}.  A preliminary account
of the analysis below was presented by \citet{lor04}.

Our goal is to produce a model of the Galactic pulsar population
which, when filtered through the above survey detection models of the
PMB and PH surveys produces a set of ``model detectable pulsars''
which closely matches the observed sample. A Monte Carlo simulation
package\footnote{\url{http://psrpop.sourceforge.net}} was developed
for this work and is freely available for checking the results
presented here, and for further use.

We simulate the Galactic pulsar distribution by drawing pulsars from
independent numerical PDFs in $R$, $z$, $L$ and $P$. The justification
for assuming no correlation between the distributions is twofold.
Firstly, no statistically significant correlations
between these parameters exist in the
known sample \citep{lor04}. Secondly, our modeling philosophy is to
create a {\it snapshot of the currently observable pulsar population},
rather than attempting to follow the time evolution \citep[see,
e.g.,][and references therein]{fk06}. Although there are advantages to
taking the time-evolution approach, particularly when investigating
the evolution of pulsar luminosities and magnetic field strengths, we
prefer to adopt the simpler procedure here which concentrates on
issues relating to the present-day spatial, period and luminosity
distributions.

We choose the four numerical PDFs as follows: for the $R$ distribution
we consider 15 equal zones between 0 and 15 kpc; for the $z$ distribution we
consider 31 equal zones between --1.5 and 1.5 kpc; for the $L$ distribution
we consider 12 logarithmically spaced zones between 0.1 and 1000
mJy~kpc$^2$ each one third of a decade wide; for the $P$ distribution
we consider 15 logarithmically spaced zones between 30~ms and 10~s
each one sixth of a decade wide. To begin with, we assume no prior
knowledge and distribute pulsars completely uniformly in space and use
flat PDFs in log $L$ and log $P$. We also experimented with other
starting PDFs and concluded that any function which reasonably
samples the parameter range produces results that are consistent
with those found below.

During the course of this work, we found that the most important
factor limiting our determination of the true spatial density of
pulsars in the Galaxy was the uncertainty in the Galactic distribution
of free electrons.  We therefore distinguish between two different
models. The first model (hereafter referred to as ``S'' for smooth),
treats the pulsar and free electron density distributions as smooth
and azimuthally symmetric functions with respect to the Galactic
centre. The second model (hereafter referred to as ``C'' for clumpy),
attempts to incorporate Galactic spiral arm structure and a
heterogeneous distribution of free electrons.

To specify the location of any model pulsar, we use a Cartesian
$(x,y,z)$ coordinate system where the Galactic centre is at the
origin, and the Sun is at (0.0,8.5,0.0)~kpc. With this definition
$R=\sqrt{x^2+y^2}$.  For a given pair of $R$ and $z$ values, then, we
can calculate the Cartesian coordinates of each model pulsar. We
achieve this in two different ways. In model S, we choose a random
azimuthal angle in the disk of the galaxy $\theta$ from a flat PDF in
the range $0 \leq \theta \leq 2\pi$.  The coordinates are then simply
$x=R \sin \theta$ and $y=R \cos \theta$. In model C, we follow the
procedure described by \citet{fk06} to weight the pulsar distribution
along model spiral arms and calculate the appropriate $x$ and $y$
positions.  In both cases, after determining the $x$, $y$ and $z$
coordinates of each pulsar, it is straightforward to compute its true
distance from the Sun $d$ and apparent Galactic coordinates, $l$ and
$b$. Finally from the distance and luminosity of each pulsar, and the
definition $L=S_{1400}d^2$, we calculate the true 1400-MHz flux
density, $S_{1400}$.

As mentioned in the previous section, the true flux density needs to
be corrected for the position offset between a pulsar and the nearest
telescope pointing in the survey.  The PMB survey comprises data from
40077 beam positions shown in Figure~\ref{fg:lb}. Although the
pointings principally cover the longitude range
$-100^{\circ}<l<50^{\circ}$ and $|b|<5^{\circ}$, as can be seen there
is some coverage of the region $-140^{\circ}<l<-100^{\circ}$ and
$|b|<3^{\circ}$.  From the sky position of each model pulsar we use
this database to find the closest beam position in the survey. To
model the PH survey, which had a total of 6456 beam positions in the
region $-140^{\circ}<l<-100^{\circ}$ and $|b|<60^{\circ}$, we randomly
choose an offset within the full-width half power point of a telescope
beam. From the corresponding position offset and true flux density, we
use Equation~\ref{equ:fluxcor} to find the effective flux density,
$S$.

To calculate DMs, we integrate two different models for the free
electron density out along the appropriate line of sight defined by
$l$ and $b$ until the distance $d$ is reached. For model S we use the
azimuthally symmetric and smooth model\footnote{We note that a more
recent electron density model proposed by \citet{gbc01} was not used
here due to its lack of predictive power for pulsars in the inner
Galaxy ($R<4$~kpc).} derived by \citet{lmt85}, while for model C, we
adopt the free electron distribution due to \citet{cl02} which
attempts to account for the clumpy nature of the interstellar medium
including spiral arm structure. As discussed by \citet{lmt85} and
\citet{cl02}, the overall random errors in both models result in
distance uncertainties of the order of 20\%.  We account for this in
both models by dithering all true distances using a Gaussian
distribution with a fractional error of 20\% and use the symbol
\,$\hat{}$\, to denote parameters affected by this distance
uncertainty.  Hence, in addition to the dithered distance, $\hat{d}$,
we have $\hat{L}=S\hat{d}^2$, $\hat{R}=\sqrt{\hat{x}^2+\hat{y}^2}$ and
$\hat{z}=\hat{d} \sin b$ as the corresponding radius, $R$, and
$z$-height estimates. For self-consistency, when comparing our model
samples with the observations below, we use the model $\hat{L}$,
$\hat{R}$ and $\hat{z}$ quantities.

Having computed the position and flux density of each pulsar, we are
now in a position to model its detectability.  To determine the model
S/N using Equation~\ref{equ:dewey}, we need to find the detected pulse
width, $W$. This can be written in terms of the intrinsic pulse width
$W_{\rm int}$ and a number of other factors in the following quadrature sum:
\begin{equation}\label{eq:weff}
W = \sqrt{W_{\rm int}^2 + W_{\rm DM}^2 + 
  t_{\rm samp}^2 + t_{\rm scatt}^2}.
\end{equation}
Here $W_{\rm DM}$ is the dispersive broadening across individual
filterbank channels, 
$t_{\rm samp}$ is the data sampling interval (multiples of 250 $\mu$s
for the PMB survey and multiples of 125 $\mu$s for the PH survey) and
$t_{\rm scatt}$ is the interstellar scattering time. In the following,
we examine each of the quantities on the right hand side of this
equation in turn.

It is well known \citep[see, e.g.,][]{lm88} that pulse widths are
correlated with pulse period.  During the course of this work, it
became obvious that the observed scatter in the $W$--$P$ plane (see
Fig.~\ref{fg:wp}) cannot be explained by a random relationship.  Some
correlation between $W$ and $P$ is to be expected, given the known
variation in beam size with pulse period \citep[see, e.g.,][]{big90b},
despite being blurred by an arbitrary viewing geometry for each
pulsar. Rather than adopt a complicated beam model, we adopt a purely
empirical approach in our simulations by assigning the intrinsic pulse
width, $W_{\rm int}$, as follows:
\begin{equation}
\label{eq:wint}
  \log W_{\rm int} = 
\log \left[ 0.06 \left(\frac{P}{\rm ms}\right)^{0.9}\right]
+ \Gamma,
\end{equation}
where $\Gamma$ is a suitably normalized Gaussian distribution about
the origin with a standard deviation of 0.3.  These parameters were
found by experimentation so that, after taking into account the
various additional pulse broadening effects described below, detected
pulse widths of the model pulsars are in reasonable agreement with the
observations (see Fig.~\ref{fg:wp}). Although the exact values of the
above numerical factors are not crucial to our final results, models
without this correlation built in were found to produce very
unsatisfactory comparisons with the real sample.

Turning now to the instrumental terms in Equation~\ref{eq:weff}, we
can insert the parameters of both surveys and write the dispersive
broadening across individual 3-MHz frequency channels as follows:
\begin{equation}
W_{\rm DM} = 9.6 \,\mu{\rm s} 
\left(\frac{\rm DM}{{\rm cm}^{-3}\,{\rm pc}}\right).
\end{equation}
As described by \citet{mlc+01}, the sampling time $t_{\rm samp}$
depends on DM$_{\rm trial}$. A file containing a list of trial DMs and
sampling times is available at
http://www.blackwellpublishing.com/products/journals/su
ppmat/MNR/..../PMSURV.DMs

Finally, in order to model interstellar scattering, we make use of the
well-known correlation between DM and the scattering timescale $t_{\rm
scatt}$ \citep[see, e.g.,][]{sdo80}. Scaling the best-fit relationship
from the recent study of \citet{bcc+04} to 1400~MHz, we find
\begin{equation}
\label{equ:tau}
{\cal T} = 0.154 {\cal D} + 1.07 {\cal D}^2 -7,
\end{equation}
where ${\cal T}$ and ${\cal D}$ are respectively the base-10
logarithms of $t_{\rm scatt}$ (in ms) and DM (in cm$^{-3}$~pc).  To
account for the dispersion about this relationship (see Fig.~4 of
Bhat et al.~2004), we dither the model scattering times using a normal
distribution around ${\cal T}$ with a standard deviation of 0.8.

The above discussion has detailed all the main steps in the simulation
which allow us to calculate a predicted S/N for each model pulsar
using Equation~\ref{equ:dewey}. Those pulsars with S/Ns greater than 9
were deemed to be theoretically detectable.  Each simulation proceeded
until 25,000 model detectable pulsars were accumulated. This number
was chosen to be much larger than the true sample size to minimize
statistical fluctuations \citep[see, e.g.,][]{bwhv92}.  For each
simulation, histograms of the observed distributions of $P$,
$\hat{L}$, $\hat{R}$ and $\hat{z}$ were formed and normalized to have
the same area as the sample of real observed pulsars.

To assess the overall quality of a simulation, we calculated the
reduced $\chi^2$ statistic summed over the bins of the $P$, $\hat{L}$,
$\hat{R}$ and $\hat{z}$ distributions. We note however that $\chi^2$
is used here only as a figure-of-merit, and not as a means for
hypothesis testing or rejection.  For a typical starting model with
flat PDFs in $R$, $L$, $z$ and $P$ the reduced $\chi^2$ was, as
expected, rather large, with typical values of order 1000. To
improve these models, for each distribution, we computed a set of
correction factors with one factor per bin defined as
\begin{equation}
  C_i = \frac{R_i-M_i}{M_i},
\end{equation}
where $R_i$ and $M_i$ are respectively the real and model observed
number of pulsars in the $i^{\rm th}$ bin of one of the $P$,
$\hat{L}$, $\hat{R}$ and $\hat{z}$ histograms. A new set of input
distributions was then generated by multiplying each bin by the
correction factor and adding this product to the original bin
value. For example, the new $z$ distribution, $z^{\prime}$, is
computed as follows:
\begin{equation}
  z^{\prime}_i = z_i + z_i \times C_i.
\end{equation}
The resulting new set of PDFs should provide a better match to the
observations. The simulation is then re-run and a new set of
correction factors computed. We found that this procedure rapidly
converged to produce a model with a much lower reduced $\chi^2$
(typically of order unity) that stabilized after only a few
iterations.  We note in passing here that there appears to be very
little covariance between the four PDFs. We found, through
experimentation, that it was not possible to reach any satisfactory
convergence by keeping one or more parameters constant during the
iterations and varying the others. In this regard, at least within the
overall limitations of the method we are using, we are confident that
the models presented below are optimal and do not represent `local
minima' solutions.

\subsection{A self-consistency test of the model}\label{sec:selfc}

To verify that the above procedure can successfully recover the
various Galactic distribution properties, we simulated two fake
populations with different radial density functions. In the first
population, pulsars were drawn from a simple Gaussian profile
\citep[see, e.g.,][]{nar87} with a standard deviation of 6.5~kpc. For
the second population we adopted the more recent profile derived by
\citet{yk04} which peaks at $\sim 4$~kpc and has a deficit of pulsars
in the inner Galaxy. In both simulations, we assumed an exponential
$z$ distribution with a scale height of 350 pc \citep[][]{mm04}, a
log-normal distribution of luminosities with a mean of --1.1 and
standard deviation 0.9 \citep[both in log 10 of mJy~kpc$^2$,][]{fk06}
and a gamma function for the spin periods which peaks at $\sim 600$~ms
as described by \citet{gh96}. The assumed pulse width versus period
relationship was given in the previous section.  Both models were run
such that 1005 model pulsars were detected by our models of the PMB
and PH surveys. The resulting model observed samples were then used as
fake observed populations, i.e.~for the purposes of this test, they
take the place of the real observed sample.

After applying the Monte Carlo procedure described in the previous
section to these fake samples, we were able to reproduce the
functional form and correct number density over the parameter space of
each model.  Fig.~\ref{fg:gy} shows the results in terms of the input
and derived radial density functions.  These tests give us confidence
that, when applied to the real observed sample of pulsars, our
procedure provides a reliable means of estimating the true spatial and
luminosity functions of pulsars in the Galaxy.

\subsection{Application to the observed pulsar sample}\label{sec:appl}

We now apply the above procedure to the real sample of 1008 isolated
pulsars detected by the PMB and PH surveys. As discussed above we
consider two different cases: a ``smooth'' (S) model of the Galaxy
using the \citet{lmt85} electron density distribution and a ``clumpy''
(C) model using the \citet{cl02} electron density distribution and
following the spiral arm modeling described by \cite{fk06}. In each
case we computed distances to both the real and model pulsars using
the appropriate electron density model.

Figs.~\ref{fg:modelS} and \ref{fg:modelC} show the real observed and
resulting model underlying distribution functions derived from our
analysis. The error bars shown here are purely statistical estimates
with a fractional uncertainty equal to $1/\sqrt{N}$, where $N$ is the
number of observed pulsars in each bin.  To parameterize our results
in a form that may be of use to others, we also attempt to fit various
analytic functions to the underlying distributions; these are shown by
the smooth curves in each figure. For the radial density profile, we
use a gamma function
\begin{equation}\label{eq:rdist}
  \rho(R) = A \left(\frac{R}{R_{\odot}}\right)^{B} 
  \exp \left(-C\left[\frac{R-R_{\odot}}{R_{\odot}}\right]\right);
\end{equation}
for the $z$ distribution, we use the exponential function
\begin{equation}\label{eq:zdist}
  N = D \exp(-|z|/E);
\end{equation}
for the luminosity distribution, we fit a simple power law 
\begin{equation} \label{eq:ldist}
  \log N = F \log L + G;
\end{equation}
finally, we fit the period
distribution to the function
\begin{equation}\label{eq:pdist}
  N(\log P) = H \exp \left(-\frac{[\log P - I]^2}{2 J^2}\right).
\end{equation}
Table~\ref{tb:fits} summarizes the resulting fit parameters $A$
through $J$ for both models. The dotted lines in the figures compare
our results with various other distributions and are described in turn
below.

\subsection{Discussion of the results}\label{sec:comp}

\subsubsection{Radial distribution}

The underlying radial density distribution of each model provides a
straightforward means of calculating the implied number of potentially
detectable pulsars (i.e., above the luminosity threshold of 0.1 mJy
kpc$^2$ and beaming toward us) in the Galaxy, $N_G$. For models S and
C respectively we find $N_G = 28000\pm 1100$ and $30000\pm 1100$.
Adopting the \citet{tm98} beaming model, these numbers translate to
$148000\pm6000$ and $155000\pm6000$ active radio pulsars in the Galaxy
with luminosities above 0.1 mJy kpc$^2$.  While these numbers are
consistent, the shape of the radial density distributions are quite
different. As seen in Figs.~\ref{fg:modelS}a and \ref{fg:modelC}a,
where the dotted lines show the radial distribution of free electrons,
we find that the pulsar distribution very closely follows that of the
electrons in both cases.  This striking correlation can be understood
when one realizes that model C has an enhancement of electrons in the
range $3<R<5$~kpc. Thus, at low Galactic longitudes, the model will
preferentially place pulsars either on the near or far side of the
Galaxy, close to the electron density enhancement. For model S, where
the electron density peaks at $R=0$, such pulsars are preferentially
placed towards the centre of the Galaxy. These different radial
distributions have implications for the detection rates of future
surveys as we discuss later in \S \ref{sec:pred}.

We conclude from the above discussion that the Galactic pulsar and
electron density distributions in the inner Galaxy are highly
covariant. Based on our current state of knowledge, i.e.~that the
\citet{cl02} model provides a far better description of the free
electron distribution than all of its predecessors, our optimal model
for the Galactic pulsar population is model C. However, given the
uncertainty in the electron distribution in the inner Galaxy, care
should be taken when interpreting the results of our fit which shows a
peak in the pulsar distribution at 3--5 kpc. Further progress on the
radial density distribution of pulsars requires independent distance
estimates for more pulsars in the inner Galactic quadrants.  Another
possibility worth exploring, though beyond the scope of this paper, is
to treat the electron density distribution as a free parameter in the
Monte Carlo simulations. For now, we simply caution that the deficit
of pulsars in the inner Galaxy noted here \citep[see also the earlier
papers by][]{joh94,yk04,lor04} depends on the assumption of a
corresponding deficit in the free electron density.

\subsubsection{$z$ distribution}\label{sec:z}

We now consider an interesting result from our fits to the $z$
distributions of both models, where it is seen that model S results in
a scale height of 330 pc compared to only 180 pc for model C. The
latter value is much lower than the canonical value of 300--350~pc
expected {\it a-priori} from an independent study of the local pulsar
population \citep{mm04} and shown by the dotted line in
Figs.~\ref{fg:modelS}b and \ref{fg:modelC}b.  As for the radial
distribution, we believe this difference to be a direct result of the
different electron density distributions used in the two models. From
a comparison of Figs.~\ref{fg:modelS}b and \ref{fg:modelC}b, we see
that the $z$ heights of the observed pulsar sample computed using the
\citet{lmt85} electron density model are much broader than when using
the \cite{cl02} model.

The discrepancy is further highlighted when one utilizes the two other
surveys carried out with the Parkes multibeam system, i.e.~the
Swinburne intermediate and high latitude pulsar surveys
\citep[respectively SIL and SHL surveys,][]{ebvb01,jac05}.  Along
with the PH survey, these Swinburne surveys sample the distribution
away from the Galactic plane and are therefore very sensitive to the
$z$ scale height of the pulsar population.  As mentioned previously,
these surveys were not included in the main analysis because the
majority of pulsars from them do not have measured flux
densities. However, our models should be able to reproduce the
relative yield of the PMB and PH surveys alongside the SIL and SHL
surveys. We have carried out such an analysis and compare the observed
pulsar samples from these surveys with the various models in
Table~\ref{tb:yields}. As can be seen, the expected survey detections
for model S are generally in good agreement with the actual survey
results. For model C, however, the predicted yield for all surveys
away from the Galactic plane is much lower than observed.

As a final sanity check, we have verified through additional
simulations similar to those described in \S \ref{sec:selfc}, that our
modeling procedure is capable of reproducing any reasonable input
scale height distribution.  Based on these results, we conclude that
there is a potential inconsistency when using the \citet{cl02} model
to derive the scale height of pulsars and consider a more likely value
to be that found by model S, i.e.~330 pc. The apparent clustering of
the PMB pulsars towards small $z$ heights when using the \citet{cl02}
model was also noted by \cite{kbm+03}. It appears that further
investigation of the $z$-distribution in the electron density model is
now warranted. While \citet{mm04} also used the \citet{cl02} electron
density model, their analysis only considered pulsars within a
cylinder of radius 3 kpc centred on the Sun. This local population
analysis was done to minimize selection effects. Although our analysis
attempts to correct for these effects as far as possible, we believe
the \citet{cl02} distance model is biasing our results towards an
artificially lower scale height.

In an attempt to correct for this bias, we have introduced a modified
version of model C, denoted by model C$^{\prime}$, which is identical
with the original model with the exception of having an exponential
$z$ distribution with a scale height of 330 pc. As can be seen from
the yields of this model in Table \ref{tb:yields}, the revised scale
height is in much better agreement with pulsar surveys carried out at
higher Galactic latitudes. We return to this model for the discussions
on pulsar current analysis and survey predictions (sections
\ref{sec:pca} and \ref{sec:pred}).

\subsubsection{Luminosity distribution}\label{sec:ldist}

For both models S and C, we find that the simple power law given in
Equation \ref{eq:ldist} provides an adequate description of the
underlying luminosity function for $L>0.1$~mJy~kpc$^2$. However, the
slope of the distribution (--0.6 for model S and --0.8 for model C) is
somewhat flatter than the canonical value $d \log N/ d \log L = -1$
derived in early studies of the pulsar population \citep[see,
e.g.,][]{tm77}.  Overlaid on both Figs.~\ref{fg:modelS}c and
\ref{fg:modelC}c is the log-normal fit to the luminosity distribution
from the optimal model of \citet{fk06}. As can be seen, given the
uncertainties involved, both models S and C are in reasonable
agreement with the results of \citet{fk06}.

The behaviour of the luminosity function below 0.1~mJy~kpc$^2$ is not
well constrained by our analysis. Although we began our study using a
luminosity distribution which extended to 0.01~mJy~kpc$^2$, we found
that the few pulsars in the observed sample with $L<0.1$~mJy~kpc$^2$
could be accounted for simply by a luminosity function with a lower
limit of 0.1~mJy~kpc$^2$ and the dithering of luminosities resulting
from a 20\% distance uncertainty. Better constraints on the shape of
the low end of the pulsar luminosity function should be possible from
an analysis of deep targeted searches for young pulsars \citep[for a
recent review, see][]{cam04}.

\subsubsection{Period distribution}

In Figs.~\ref{fg:modelS}d and \ref{fg:modelC}d, we compare our period
distribution with the parent distribution favoured by \citet{kgk04} in
their analysis of pulse-width statistics. As can be seen, our
distribution is significantly different, with a much higher proportion
of short-period pulsars. Given the effort to quantify and
account for period-dependent selection effects in this paper, we
believe our fit using the log-normal distribution given in Equation
\ref{eq:pdist} provides a much better description of the parent
population of normal pulsars and recommend its adoption in future
pulse-width analyses.

\subsection{Pulsar current analysis}\label{sec:pca}

To compute the birth rate of the population, we follow previous
authors \citep{pb81,vn81,lbdh93,lml+98,vml+04} and perform a pulsar
current analysis. We closely follow the approach described in detail
by \citet{vml+04} and calculate a weight or ``scale factor'' for each
of the 1008 pulsars in our main sample. The analysis proceeds by
placing each of these pulsars in every position of our model galaxy
and recording the number of detections made by our model PMB and PH
surveys. The model galaxy used for these calculations was model
C$^{\prime}$, i.e.~the modified version of model C which has a $z$
scale height of 330 pc and produces the best match to the other survey
detection rates (see \S \ref{sec:z}).  To compute the apparent flux
density, we use the observed luminosity of each pulsar and apply the
empirical relationship given in Equation \ref{eq:wint} to calculate an
intrinsic pulse width at each position in the model Galaxy. The scale
factor is then the ratio of the total number of positions tried to the
number of detections found.

Having obtained a set of scale factors, the current or flow of pulsars
from short to longer periods can be calculated by binning the sample
into period intervals of width $\Delta P$.  For a given period bin
containing $n_{\rm psr}$ pulsars, following \citet{vn81}, we define
the pulsar current
\begin{equation}
  J(P) = \frac{1}{\Delta P} \left( \sum_{i=1}^{n_{\rm psr}} 
  \frac{\xi_i \dot{P}_i}{f_i} \right),
\end{equation}
where $\xi_i$ is the scale factor of the $i^{\rm th}$ pulsar, with
period derivative $\dot{P}_i$ and $f_i$ is the fraction of $4 \pi$ sr
of the sky covered by the radio beam. In the discussion below, we
quote results for `potentially observable pulsars' (i.e.~$f_i=1$) and
assuming various beaming models for which $f_i<1$. The uncertainty in
the current for each period bin is dominated by the statistical error
in the scale factors. These are calculated, again following
\citet{vn81}, as the square root of the sum of the squares of the
scale factors in each period bin.

A histogram of pulsar current versus period is shown for the
population as a whole in Fig.~\ref{fg:current} from which we infer a
birth rate (i.e.~maximum current value) of potentially observable
pulsars with $L>0.1$~mJy~kpc$^2$ to be $0.34\pm 0.05$~pulsars per
century. As in earlier analysis, we find here no requirement for the
birth (or ``injection'') of pulsars with periods $\sim
0.5$~s. Following these authors, we have also examined the birth rate
contribution as a function of inferred dipole magnetic field strength,
$B$. Using the same grouping criteria as \citet{vml+04}, we find
birth rates of $0.04\pm0.02$, $0.14\pm0.07$ and $0.16\pm0.04$~pulsars
per century respectively for the low ($B<9 \times 10^{11}$~G), medium
($9 \times 10^{11} \leq B \leq 2.5 \times 10^{12}$~G) and high ($B>
2.5 \times 10^{12}$~G) field populations.  These results agree with
the findings of \citet{vml+04}.

In Table \ref{tb:br} we use various beaming models to apply a
period-dependent correction to the pulsar current which accounts for
those pulsars whose radiation beams do not intersect with our line of
sight. We also break down these results for three different luminosity
cut-offs: 0.1, 1 and 10 mJy~kpc$^2$.  For the three most recently
published beaming models considered \citep{lm88,big90b,tm98} the
results are consistent within the uncertainties derived from the scale
factor statistics. For the earlier \citet{nv83} model, in which the
radio beams are elongated in the meridional direction, the birth rates
are systematically lower. Regardless of the beaming model adopted,
this compilation suggests that the majority ($\sim 80$\%) of all
pulsars {\it appear} to be born with luminosities
$L>1$~mJy~kpc$^2$. It should be stressed, however, that the pulsar
sample from which these results were obtained do not include the
recent discoveries made in deep targeted searches of supernova
remnants and X-ray point sources \citep[][and references
therein]{cam04}.  As was the case for the luminosity function
discussed in \S \ref{sec:ldist}, further insights into the birth
luminosities of pulsars will await a statistical analysis of these
discoveries from the targeted searches.

\subsection{Predictions for other pulsar surveys}\label{sec:pred}

We have shown in \S \ref{sec:z} that our models have generally
provided a good match to the yields of existing pulsar surveys. It is
therefore natural to apply these models to make some predictions of
the number of pulsars expected in on-going surveys. In Table
\ref{tb:yields} we list the number of expected pulsar detections for
two such projects: the Perseus Arm (PA) survey being carried out at
Parkes and PALFA, a pulsar survey with the 7-beam Arecibo L-band Feed
Array (ALFA) system \citep{cfl+06,lsf+06}. Note that the predictions
below refer to normal pulsars only.

The PA survey is extending the Galactic plane coverage of
the PMB survey to the region $-160^{\circ} < l < -100^{\circ}$ and
$|b|<5^{\circ}$ where there are currently 17 non-recycled
pulsars. With the exception of a data sampling interval of 125$\mu$s,
the PA survey uses identical observing parameters and data acquisition
and processing schemes to those implemented in PMB survey.  Since the
PA survey is targeting pulsars outside the Solar circle, the expected
yield is strongly dependent on the radial distribution in the outer
Galaxy. This can be seen by comparing the number of detections for
model S (62) and C$^{\prime}$ (32).  Upon completion, it is expected
that this survey will provide excellent constraints on the radial
distribution for $R>8.5$~kpc.

The PALFA survey is currently observing two regions of the Galactic
plane visible from Arecibo with $|b|<5^{\circ}$ at an observing
frequency of 1.4~GHz. The inner Galaxy survey (PALFAi) covers the
region $32^{\circ}<l<77^{\circ}$ while the anticentre survey (PALFAa)
covers the region $168^{\circ}<l<214^{\circ}$.  A detailed description
of the survey parameters and initial discoveries can be found in
\cite{cfl+06}.  Prior to these surveys, the number of normal pulsars
known in these ranges was 200 for the inner Galaxy and 6 for the
anticentre. Models S and C therefore predict of order 350 new
discoveries in the inner Galaxy. While this is significantly lower
than estimates made by \cite{cfl+06}, it is consistent with recent
independent work carried out by Faucher-Gigu\`ere \& Kaspi (private
communication).  As for the PA survey, the lower expected PALFA yields
in the anticentre region (of order 50 or 25 discoveries for models S
and C) depend sensitively on the radial density function in the outer
Galaxy.

\section{CONCLUSIONS}\label{sec:conc}

We have presented discovery and follow-up parameters for 142 pulsars
discovered in the Parkes multibeam survey. Using a sample of 1008
normal pulsars from this survey and the Parkes high latitude survey we
have carried out Monte Carlo simulations to investigate various
aspects of the Galactic population.  Our main conclusions are as
follows:

\begin{itemize}

\item The derived Galactic distribution of pulsars and assumed
distribution of free electrons in the Galaxy are strongly
coupled. Based on the electron density model of \citet{cl02}, we infer
a radial distribution for the pulsar population which peaks at 4 kpc
from the Galactic centre. Although further surveys at higher
frequencies will be useful, it is more important to have independent
constraints on the distances to currently known pulsars in this region
to confirm the proposed radial distribution.

\item The $z$ distribution as inferred from our simulations is also
strongly influenced by the electron density model used. Using the
\citet{cl02} model in our simulations, the best-fitting model
population under-predicts the number of pulsars found in the Swinburne
intermediate and high latitude surveys. To overcome this bias, we have
increased the scale height of the model population to 330 pc. We
believe that this apparent inconsistency will provide a very useful
constraint for future revisions of the Galactic distribution of free
electrons.

\item The luminosity distribution we derive follows a simple power law
with a slope of $d \log N/d \log L \sim -0.8$. While this is somewhat
flatter than previous estimates, our luminosity function is consistent
with that derived independently by \citet{fk06}.  The shape of the
luminosity function below 0.1~mJy kpc$^2$ is not well constrained by
our analysis. Although 6 pulsars with $L<0.1$~mJy~kpc$^2$ are known in
our current sample, their existence can be explained by distance
uncertainties and a luminosity function with a lower bound at 0.1~mJy
kpc$^2$. We stress again here that an analysis of recent discoveries
of faint young pulsars in targeted searches will provide further
constraints on the luminosity function of pulsars below 0.1
mJy~kpc$^2$.

\item Applying a pulsar current analysis to the observed sample and a
modified version of model C, we find the birth rate of the potentially
observable population to be $0.34\pm 0.05$ pulsars per century.
Applying the \cite{tm98} beaming model to account for unbeamed
pulsars, we find a birth rate of $1.4\pm 0.2$ pulsars per century for
luminosities above 0.1~mJy kpc$^2$. There is no evidence for any
injection of pulsars with birth periods $\sim 0.5$~s into the
population. We agree with the findings of \citet{vml+04} that most of
the birth rate comes from high magnetic field pulsars.

\item Predictions for current surveys of the outer Galaxy at Parkes
and Arecibo depend sensitively on the form of the radial distribution
outside the Solar circle. Although this distribution is poorly known
at present these surveys are expected to provide further constraints
in the near future.  The Arecibo multibeam survey of the inner Galaxy
is expected to yield of order 350 new discoveries.  We expect 
these surveys to provide further constraints on the population
parameters derived in this work.

\end{itemize}

The Parkes multibeam surveys provide a superb database with which to
study various aspects of the pulsar population.  The analysis
presented here has attempted to be as straightforward as possible with
the minimum of assumptions made. Further more detailed studies such as
presented recently by \citet{fk06} are warranted and will certainly
provide further insights into the underlying population properties
discussed here.

\section*{Acknowledgments} 

We thank Simon Johnston for useful discussions concerning the radial
density distribution of pulsars and Claude-Andr\`e Faucher-Gigu\`ere
for comments on an earlier version of this manuscript.  We gratefully
acknowledge technical assistance with hardware and software provided
by Jodrell Bank Observatory, CSIRO ATNF, Osservatorio Astronomico di
Cagliari and the Swinburne centre for Astrophysics and Supercomputing.
The Parkes radio telescope is part of the Australia Telescope which is
funded by the Commonwealth of Australia for operation as a National
Facility managed by CSIRO.  The Arecibo Observatory, a facility of the
National Astronomy and Ionosphere Center, is operated by Cornell
University under a cooperative agreement with the U.S. National
Science Foundation.  DRL was a University Research Fellow funded by
the Royal Society for the duration of most of this work.  IHS holds an
NSERC UFA and is supported by a Discovery Grant.  FC acknowledges
support from NSF grant AST-05-07376. NDA, AP and MB received support
from the Italian Ministry of University and Research under the
national program PRIN-MIUR 2005. This work made use of the facilities 
of the ATNF Pulsar Catalogue.

\bibliographystyle{mn2e}

\begin{figure*} 
\centerline{\psfig{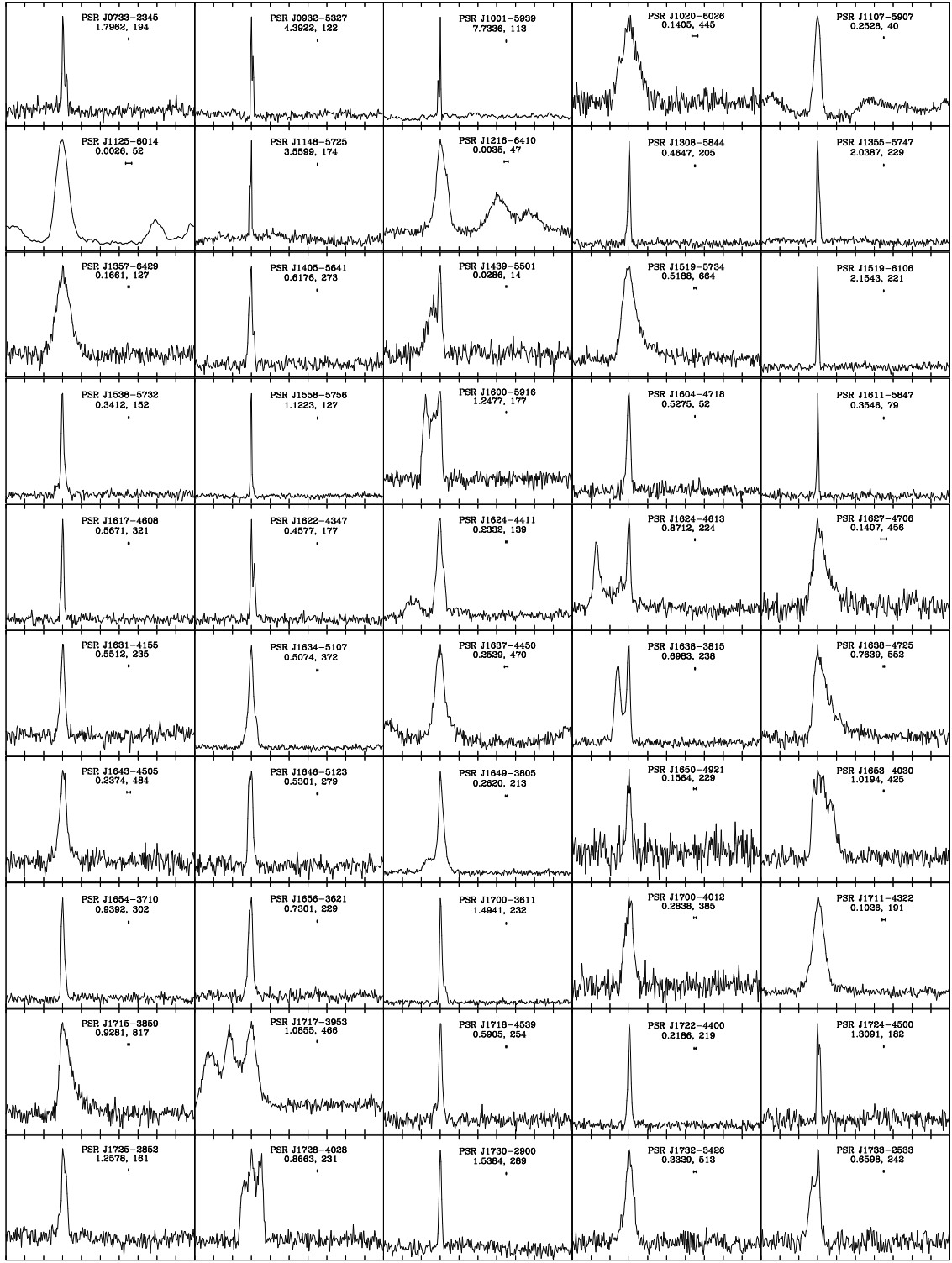}} 
\caption{Mean 1400\,MHz pulse profiles for 142 pulsars discovered in
this phase of the multibeam survey. The highest point in the profile
is placed at phase 0.3. Profiles have been compensated for the effects
of the high-pass filter in the digitization system \citep{mlc+01}. For
each profile, the pulsar Jname, pulse period (s) and dispersion
measure (cm$^{-3}$~pc) are given. The small horizontal bar under the
period indicates the effective resolution of the profile by adding the
bin size to the effects of interstellar dispersion in quadrature.}
\label{fg:prf}
\end{figure*}
\addtocounter{figure}{-1}
\begin{figure*} 
\centerline{\psfig{file=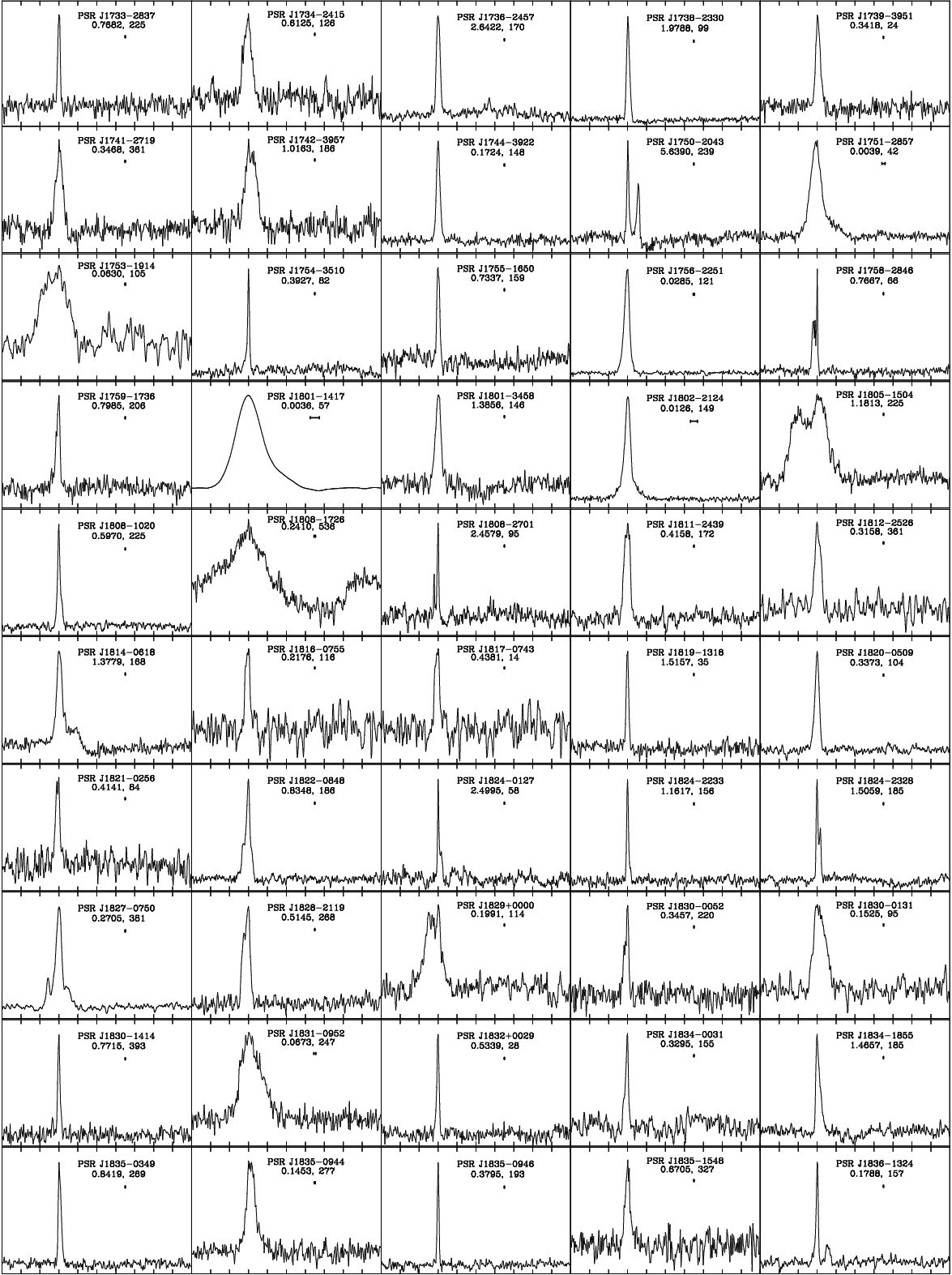,width=150mm}} 
\caption{-- {\it continued}}
\end{figure*}
\addtocounter{figure}{-1}
\begin{figure*}
\centerline{\psfig{file=plot3.ps,width=150mm}} 
\caption{-- {\it continued}}
\end{figure*}

\begin{figure*}
\centerline{\psfig{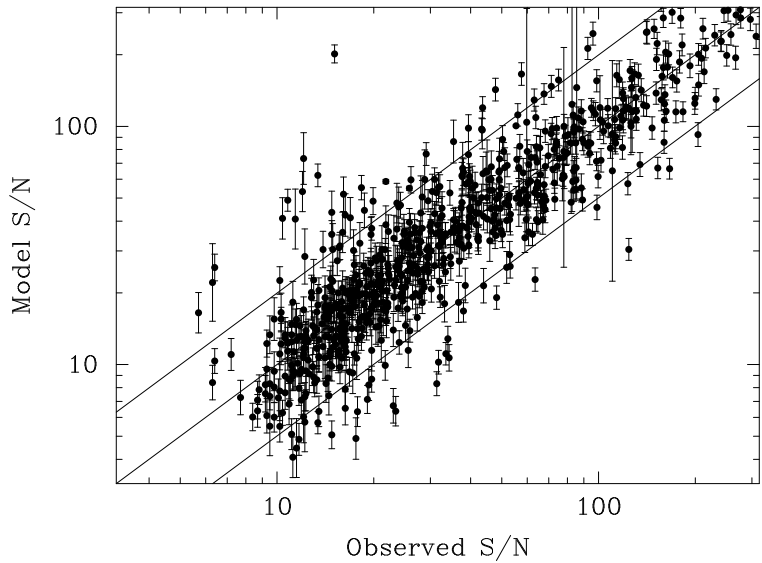}} 
\caption{Model versus observed S/N ratios calculated as described in
Section \ref{sec:sdmodel}. The central diagonal line shows equality
between theory and observation, while the outer lines show the range
of a factor of two. Uncertainties in the model S/N values are calculated
directly from the uncertainties in the observed flux densities.}
\label{fg:modelsn}
\end{figure*}

\begin{figure*}
\centerline{\psfig{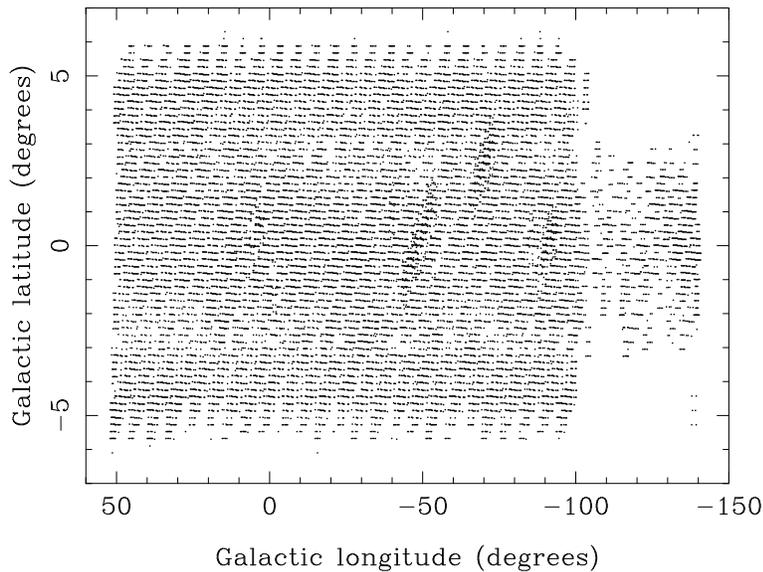}} 
\caption{Distribution of observed beam positions in Galactic
coordinates.  Note that, for clarity, the size of each dot is much
smaller than a beam width. The positions are spaced by the full-width
half maximum and form an essentially uniform coverage of the region
$-100^{\circ}<l<50^{\circ}$.  At lower longitudes, i.e.~outside the
main survey region, the coverage is incomplete. A file containing the
exact sky positions shown here is available at
http://www.blackwellpublishing.com/products/journals/suppmat/MNR/..../PMSURV.beams}
\label{fg:lb}
\end{figure*}

\begin{figure*}
\centerline{\psfig{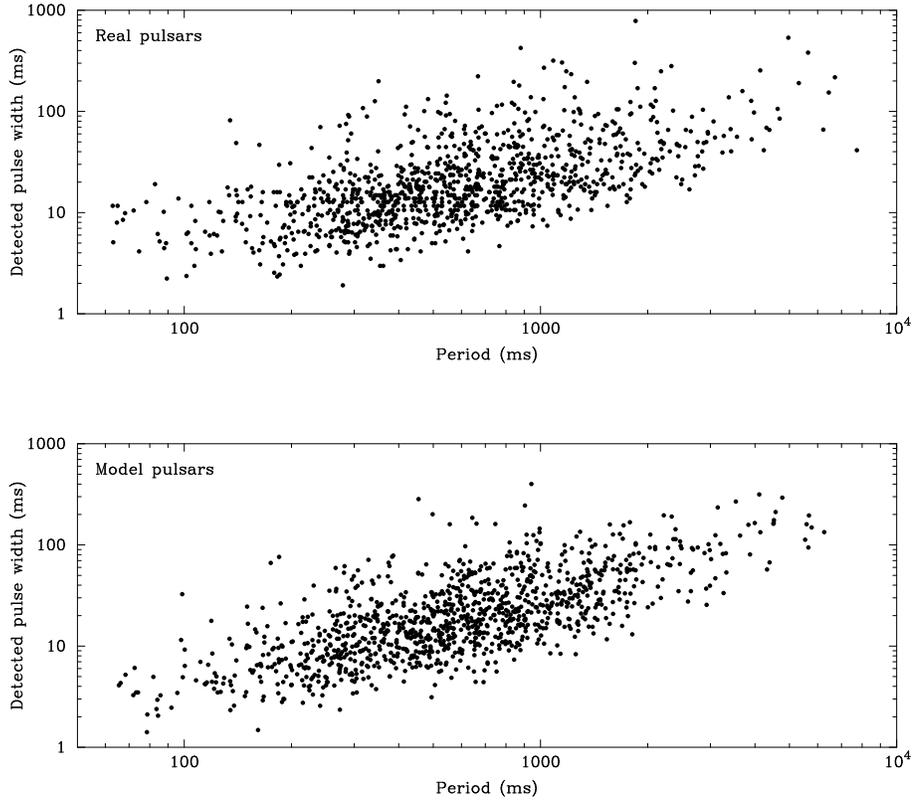}} 
\caption{Distribution of observed pulse widths as a function of pulse
period for the observed sample (top) and a model sample (bottom). The
model pulsars were generated assuming an empirically derived
correlation given in Equation \ref{eq:wint}.}
\label{fg:wp}
\end{figure*}

\begin{figure*}
\centerline{\psfig{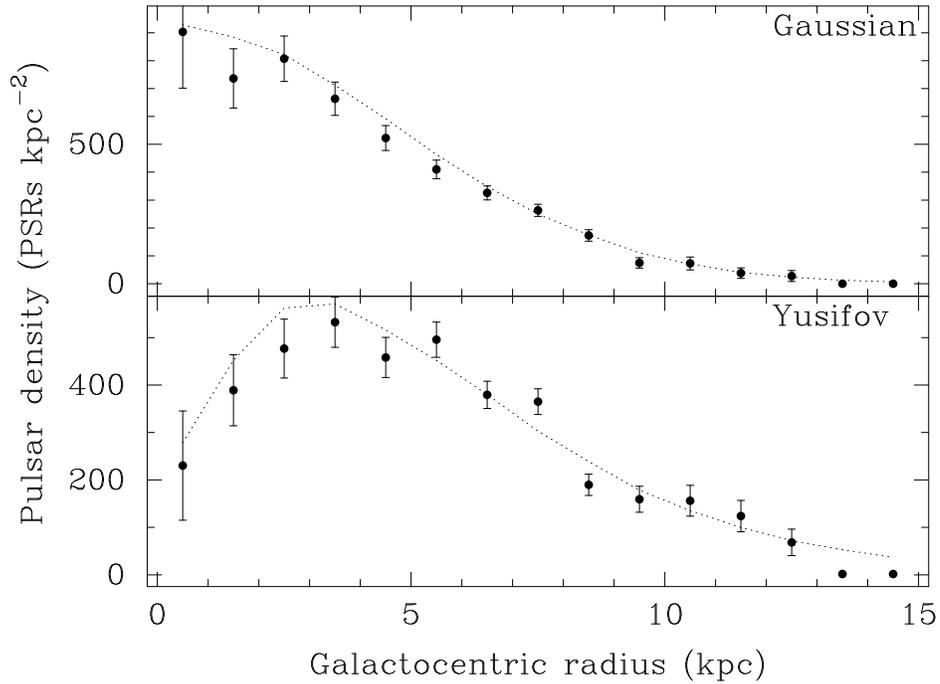}} 
\caption{Underlying (dotted lines) and derived (points) radial density
functions run for two tests of the Galactic distribution
determination. The top panel shows an input population with a Gaussian
radial profile, while the lower panel shows a population generated
using Yusifov \& K\"u\c c\"uk's (2004) radial density profile (see
section \ref{sec:selfc}).}
\label{fg:gy}
\end{figure*}

\begin{figure*}
\centerline{\psfig{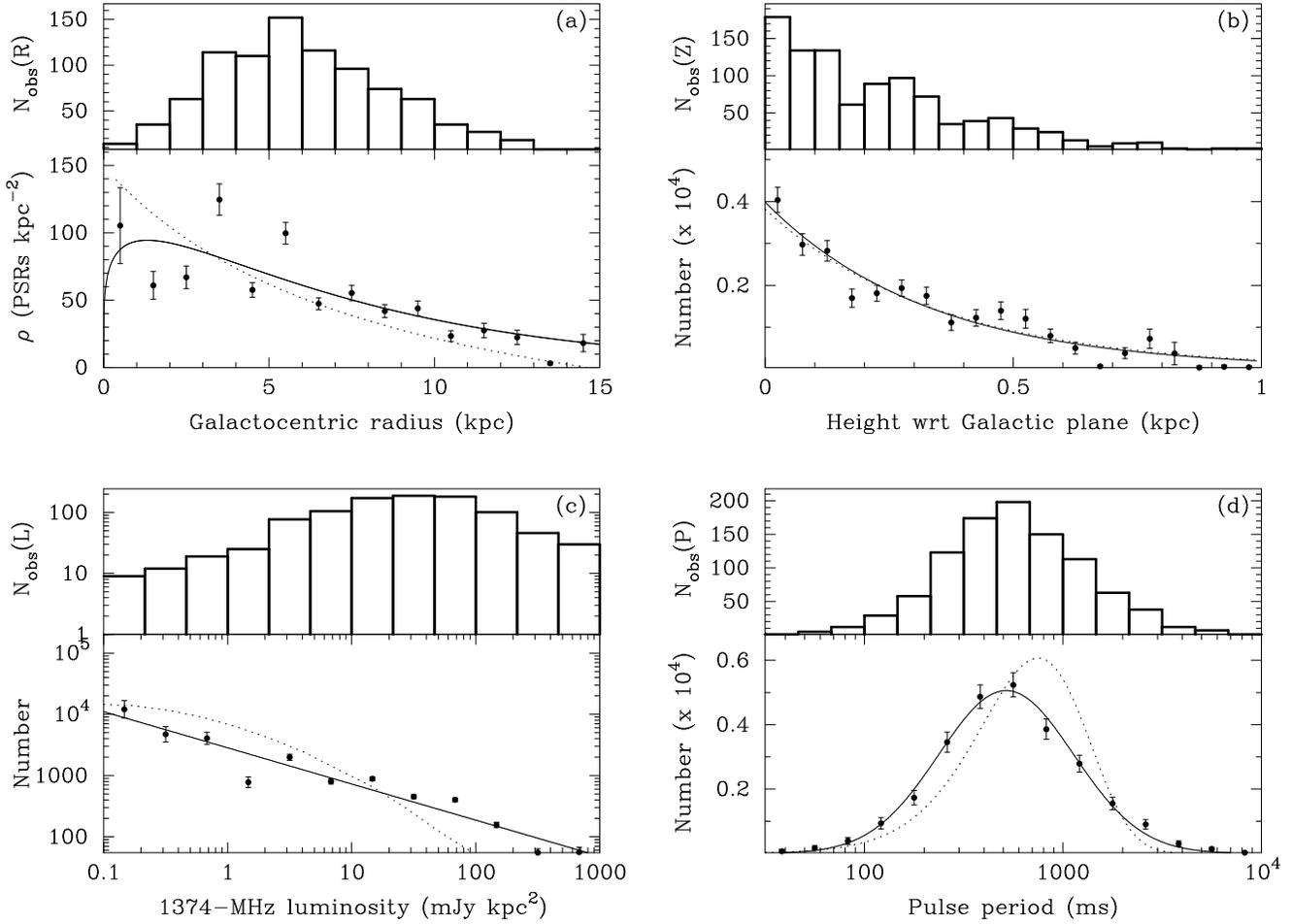}} 
\caption{Observed number distribution from our input sample (upper
panels) and derived distributions for model S (lower panels) for the
parameters: (a) $\rho(R)$; (b) $z$; (c) $L$; (d) $P$. The solid curves
are the smooth analytic functions fitted to the data (see \S
\ref{sec:appl}).  The dotted curves show: (a) the assumed radial density
function of free electrons used; (b) an exponential $z$ distribution
with a scale height of 350 pc; (c) a log-normal fit to the optimal
pulsar population model derived by \citet{fk06}; (d) a parent period
distribution used by \citet{kgk04} in a study of pulse-width statistics.}
\label{fg:modelS}
\end{figure*}

\begin{figure*}
\centerline{\psfig{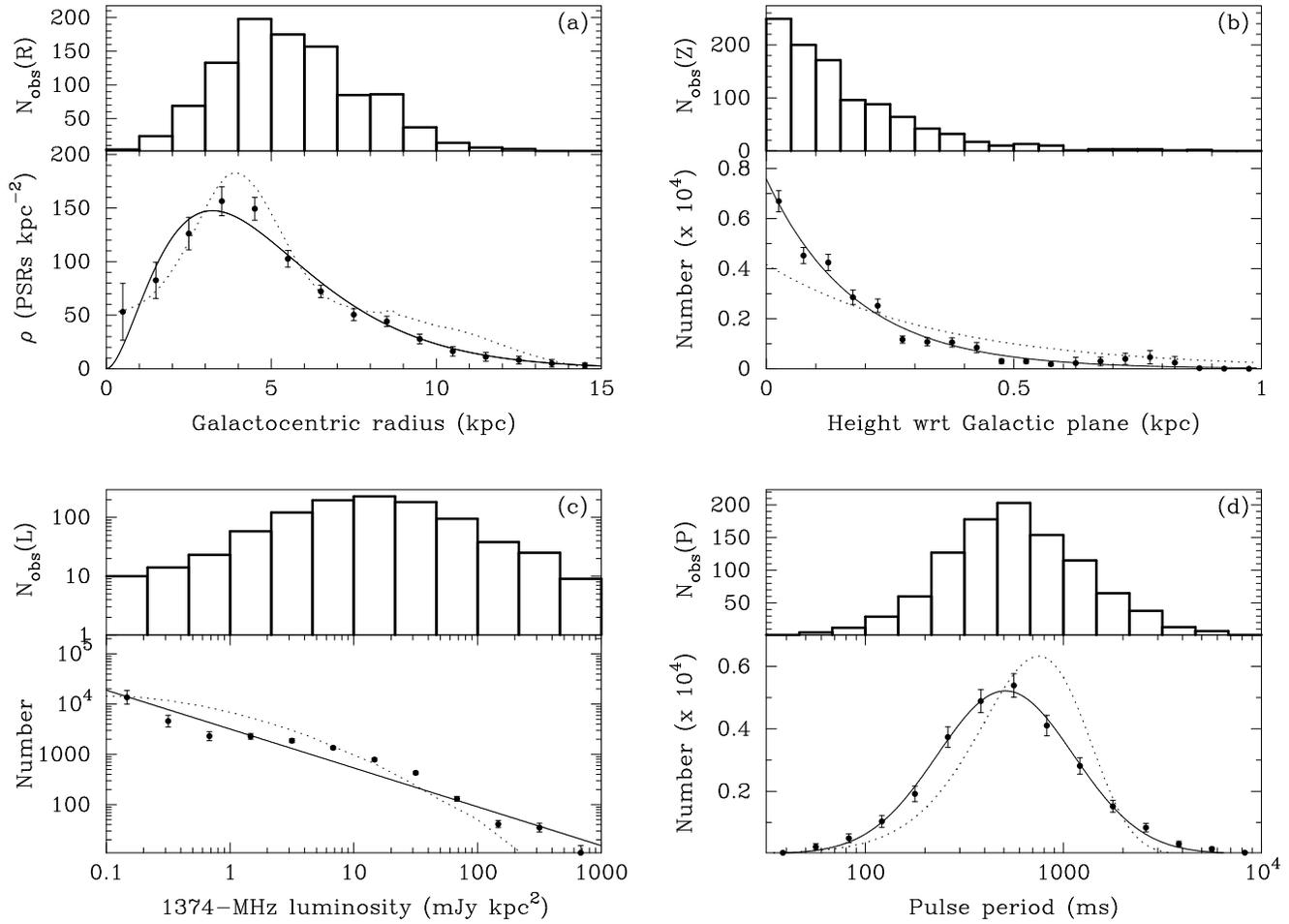}} 
\caption{Observed number distribution from our input sample (top
panels) and derived distributions for model C (lower panels) for the
parameters: (a) $\rho(R)$; (b) $z$; (c) $L$; (d) $P$. Solid and dotted
curves are as described in the caption to Fig.~\ref{fg:modelS}.}
\label{fg:modelC}
\end{figure*}

\begin{figure*}
\centerline{\psfig{file=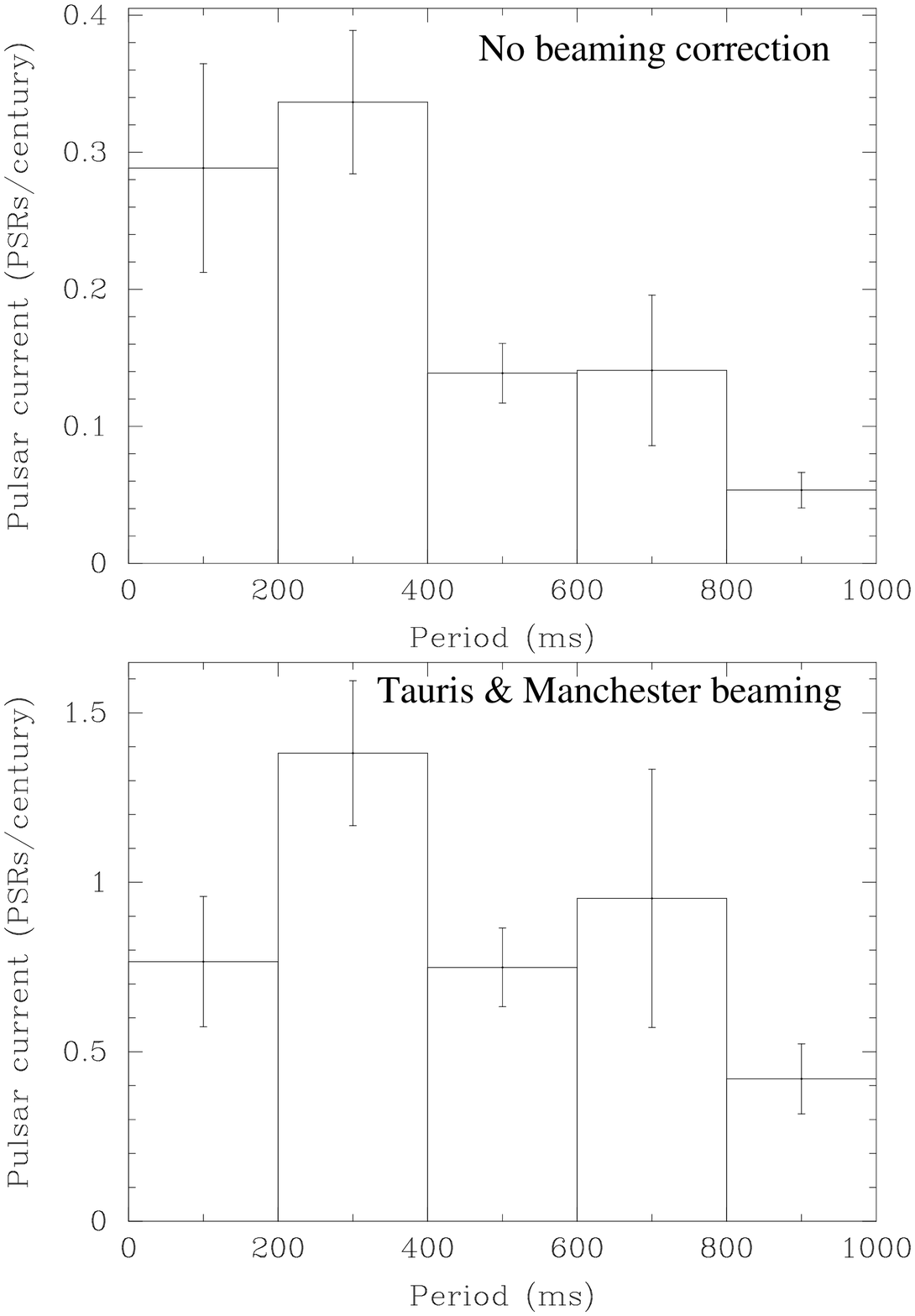,width=175mm,angle=0}} 
\caption{Histograms of pulsar current as a function of period showing
the distribution for potentially observable pulsars (top) and after
applying the beaming model of \citet[][below]{tm98} with
luminosities greater than 0.1 mJy~kpc$^2$. The error bars shown are
statistical estimates based on the sum of the squares of the
individual scale factors in each bin \citep{vn81}.}
\label{fg:current}
\end{figure*}

\clearpage
\begin{table*} 
\caption{Positions, flux densities and pulse widths for 142 pulsars
discovered in this phase of the Parkes multibeam pulsar survey. Radial
angular distances are given in units of beam radii. Pulse widths at 10
percent of the peak are given only for high signal--to--noise
profiles.}
\label{tb:posn}
\begin{center}\begin{footnotesize}
\begin{tabular}{lllrrrcrlrr}
\hline
PSR J & R.A. (J2000) & Dec. (J2000) & \multicolumn{1}{c}{$l$} & \multicolumn{1}{c}{$b$} & Beam & Radial & S/N & $S_{1400}$ & W$_{50}$ & W$_{10}$ \\
 & (h~~~m~~~s) & (\degr ~~~\arcmin ~~~\arcsec) & \multicolumn{1}{c}{(\degr)} &\multicolumn{1}{c}{(\degr)} & & Distance & & (mJy) & (ms) & (ms) \\ 
\hline 
0733$-$2345 & 07:33:24.60(4) & $-$23:45:56.2(11) & 238.71 & $-2.03$ & 4 & 0.50 & 9.0 & 0.07 & 19 & 63\\
0932$-$5327 & 09:32:22.86(14) & $-$53:27:11.0(16) & 275.89 & $-1.38$ & 8 & 0.35 & 25.1 & 0.15 & 59 & 93\\
1001$-$5939 & 10:01:32.23(14) & $-$59:39:17.8(7) & 282.94 & $-3.55$ & 7 & 0.81 & 111.2 & 0.15 & 39 & 160\\
1020$-$6026 & 10:20:11.40(7) & $-$60:26:06.8(5) & 285.30 & $-2.85$ & 12 & 0.52 & 12.6 & 0.14 & 14 & 26\\
1107$-$5907 & 11:07:34.46(4) & $-$59:07:18.7(3) & 289.94 & $1.11$ & 4 & 0.90 & 14.8 & 0.18 & 9.3 & 170\\
\\
1125$-$6014 & 11:25:55.2180(4) & $-$60:14:06.608(4) & 292.50 & $0.89$ & 12 & 0.80 & 9.3 & 0.05 & 0.21 & 1.8\\
1148$-$5725 & 11:48:28.53(19) & $-$57:25:12.6(17) & 294.51 & $4.42$ & 3 & 0.82 & 18.2 & 0.12 & 51 & 68\\
1216$-$6410 & 12:16:07.3396(9) & $-$64:10:09.226(8) & 299.10 & $-1.56$ & 2 & 0.50 & 15.7 & 0.05 & 0.22 & 3\\
1308$-$5844 & 13:08:21.08(3) & $-$58:44:13.8(4) & 305.13 & $4.06$ & 6 & 0.38 & 54.7 & 0.21 & 5.9 & 15\\
1355$-$5747 & 13:55:36.95(15) & $-$57:47:15(3) & 311.43 & $4.02$ & 6 & 0.42 & 46.2 & 0.36 & 29 & 57\\
\\
1357$-$6429 & 13:57:02.43(2) & $-$64:29:30.2(1) & 309.92 & $-2.51$ & 2 & 0.96 & 34.2 & 0.44 & 15 & 31\\
1405$-$5641 & 14:05:12.41(4) & $-$56:41:24.2(7) & 312.97 & $4.74$ & 3 & 0.64 & 15.1 & 0.10 & 12 & 29\\
1439$-$5501 & 14:39:39.7420(4) & $-$55:01:23.621(6) & 318.10 & $4.63$ & 9 & 0.34 & 17.3 & 0.42 & 1.7 & 3.4\\
1519$-$5734 & 15:19:47.59(6) & $-$57:34:13.6(7) & 321.80 & $-0.24$ & 9 & 0.74 & 24.0 & 0.45 & 40 & 82\\
1519$-$6106 & 15:19:35.81(11) & $-$61:06:54.6(12) & 319.88 & $-3.22$ & 4 & 0.01 & 38.3 & 0.19 & 25 & 47\\
\\
1538$-$5732 & 15:38:18.18(3) & $-$57:32:29.3(4) & 323.87 & $-1.62$ & 1 & 0.28 & 20.1 & 0.35 & 5 & 13\\
1558$-$5756 & 15:58:30.53(3) & $-$57:56:26.1(4) & 325.73 & $-3.60$ & 1 & 0.38 & 63.1 & 0.19 & 9.5 & 19\\
1600$-$5916 & 16:00:35.0(4) & $-$59:16:59(3) & 325.06 & $-4.80$ & 5 & 0.54 & 24.9 & 0.33 & 127 & 146\\
1604$-$4718 & 16:04:29.41(4) & $-$47:18:47.9(6) & 333.40 & $3.81$ & 4 & 0.87 & 18.5 & 0.22 & 11 & 23\\
1611$-$5847 & 16:11:51.313(6) & $-$58:47:42.33(9) & 326.46 & $-5.41$ & 6 & 0.75 & 53.1 & 0.11 & 2.8 & 6.5\\
\\
1617$-$4608 & 16:17:35.105(13) & $-$46:08:36.5(3) & 335.84 & $3.14$ & 1 & 0.92 & 20.7 & 0.15 & 7.7 & 17\\
1622$-$4347 & 16:22:30.060(18) & $-$43:47:21.4(3) & 338.12 & $4.20$ & 13 & 0.57 & 19.1 & 0.18 & 9.8 & 17\\
1624$-$4411 & 16:24:21.361(18) & $-$44:11:33.8(5) & 338.07 & $3.68$ & 10 & 0.56 & 39.5 & 0.48 & 8.1 & 50\\
1624$-$4613 & 16:24:18.7(3) & $-$46:13:02(9) & 336.62 & $2.26$ & 10 & 0.56 & --- & 0.39 & 168 & 197\\
1627$-$4706 & 16:27:28.75(4) & $-$47:06:50.2(7) & 336.35 & $1.25$ & 1 & 0.76 & 9.1 & 0.10 & 9 & 23\\
\\
1631$-$4155 & 16:31:18.33(7) & $-$41:55:03.7(20) & 340.60 & $4.33$ & 9 & 0.78 & 23.6 & 0.19 & 14 & 28\\
1634$-$5107 & 16:34:04.91(11) & $-$51:07:44.8(15) & 334.17 & $-2.29$ & 7 & 0.77 & 38.7 & 0.35 & 14 & 39\\
1637$-$4450 & 16:37:53.15(3) & $-$44:50:26.3(8) & 339.25 & $1.48$ & 12 & 1.15 & 15.4 & 0.40 & 14 & 107\\
1638$-$3815 & 16:38:37.349(18) & $-$38:15:03.4(7) & 344.26 & $5.76$ & 3 & 0.99 & 28.0 & 0.62 & 57 & 74\\
1638$-$4725 & 16:38:12.94(11) & $-$47:25:32(3) & 337.36 & $-0.30$ & 5 & 1.41 & 14.9 & 0.32 & 52 & 132\\
\\
1643$-$4505 & 16:43:36.97(9) & $-$45:05:46.0(16) & 339.73 & $0.55$ & 1 & 0.92 & 17.3 & 0.28 & 9 & 19\\
1646$-$5123 & 16:46:36.35(4) & $-$51:23:14.6(5) & 335.28 & $-3.93$ & 3 & 0.81 & 20.9 & 0.17 & 15 & 24\\
1649$-$3805 & 16:49:48.222(19) & $-$38:05:59.1(3) & 345.82 & $4.19$ & 13 & 0.56 & 61.9 & 1.0 & 8 & 34\\
1650$-$4921 & 16:50:35.100(6) & $-$49:21:03.88(12) & 337.25 & $-3.11$ & 9 & 1.51 & 15.9 & 0.16 & 3.8 & --\\
1653$-$4030 & 16:53:34.23(14) & $-$40:30:01.4(48) & 344.42 & $2.11$ & 12 & 0.88 & 16.2 & 0.40 & 122 & 161\\
\\
1654$-$3710 & 16:54:44.426(19) & $-$37:10:57.1(9) & 347.15 & $4.02$ & 13 & 0.14 & 31.3 & 0.22 & 18 & 35\\
1656$-$3621 & 16:56:32.955(15) & $-$36:21:59.7(6) & 348.01 & $4.25$ & 6 & 0.22 & 42.9 & 0.29 & 20 & 40\\
1700$-$3611 & 17:00:49.32(5) & $-$36:11:53(3) & 348.68 & $3.67$ & 12 & 0.82 & 46.9 & 0.51 & 22 & 60\\
1700$-$4012 & 17:00:38.55(4) & $-$40:12:38.6(12) & 345.49 & $1.23$ & 11 & 0.59 & 10.2 & 0.13 & 16 & 27\\
1711$-$4322 & 17:11:10.561(12) & $-$43:22:53.1(5) & 344.14 & $-2.24$ & 12 & 0.79 & 11.1 & 0.26 & 7 & 16\\
\\
1715$-$3859 & 17:15:37.89(3) & $-$38:59:25.1(9) & 348.19 & $-0.35$ & 12 & 0.80 & 15.0 & 0.54 & 61 & 128\\
1717$-$3953 & 17:17:25.55(20) & $-$39:53:55(8) & 347.65 & $-1.16$ & 3 & 0.42 & 26.5 & 0.78 & 299 & 367\\
1718$-$4539 & 17:18:11.90(4) & $-$45:39:15.9(9) & 343.02 & $-4.58$ & 4 & 0.86 & 12.9 & 0.08 & 11 & 36\\
1722$-$4400 & 17:22:46.517(12) & $-$44:00:33.2(3) & 344.84 & $-4.32$ & 12 & 0.77 & 17.3 & 0.22 & 4 & 8\\
1724$-$4500 & 17:24:25.8(3) & $-$45:00:15(4) & 344.18 & $-5.12$ & 1 & 0.83 & 12.3 & 0.05 & 28 & 38\\
\\
1725$-$2852 & 17:25:09.9(3) & $-$28:52:22(17) & 357.64 & $3.80$ & 5 & 0.66 & 15.3 & 0.25 & 37 & 79\\
1728$-$4028 & 17:28:27.68(17) & $-$40:28:10(4) & 348.38 & $-3.22$ & 1 & 0.38 & 76.2 & 0.69 & 102 & 120\\
1730$-$2900 & 17:30:08.28(10) & $-$29:00:46(12) & 358.12 & $2.82$ & 11 & 0.21 & 21.4 & 0.13 & 23 & 44\\
1732$-$3426 & 17:32:07.18(7) & $-$34:26:05(3) & 353.82 & $-0.51$ & 8 & 0.51 & 13.5 & 0.24 & 14 & 33\\
1733$-$2533 & 17:33:25.86(8) & $-$25:33:11(18) & 1.43 & $4.09$ & 1 & 0.49 & 11.9 & 0.10 & 35 & 57\\
\\

\hline 
\end{tabular}\end{footnotesize}\end{center}\end{table*}
\addtocounter{table}{-1}\begin{table*} 
\begin{center}\begin{footnotesize}
\caption{-- {\it continued}}
\begin{tabular}{lllrrccrlrr}
\hline
PSR J & R.A. (J2000) & Dec. (J2000) & \multicolumn{1}{c}{$l$} & \multicolumn{1}{c}{$b$} & Beam & Radial & S/N & $S_{1400}$ & W$_{50}$ & W$_{10}$ \\
 & (h~~~m~~~s) & (\degr ~~~\arcmin ~~~\arcsec) & \multicolumn{1}{c}{(\degr)} &\multicolumn{1}{c}{(\degr)} & & Distance & & (mJy) & (ms) & (ms) \\ 
\hline 
1733$-$2837 & 17:33:33.90(5) & $-$28:37:33(8) & 358.86 & $2.40$ & 10 & 0.24 & 14.2 & 0.07 & 12 & 20\\
1734$-$2415 & 17:34:41.6(3) & $-$24:15:20(90) & 2.68 & $4.55$ & 13 & 0.49 & 23.9 & 0.29 & 31 & 51\\
1736$-$2457 & 17:36:45.44(8) & $-$24:57:50(22) & 2.33 & $3.78$ & 9 & 0.79 & 64.0 & 0.52 & 62 & 103\\
1738$-$2330 & 17:38:08.8(3) & $-$23:30:47(103) & 3.73 & $4.28$ & 11 & 0.68 & 21.0 & 0.48 & 35 & 69\\
1739$-$3951 & 17:39:38.25(4) & $-$39:52:00.3(8) & 350.04 & $-4.69$ & 3 & 4.45 & 11.8 & 0.12 & 9 & 19\\
\\
1741$-$2719 & 17:41:35.06(4) & $-$27:19:23(7) & 0.91 & $1.60$ & 7 & 0.39 & 23.7 & 0.20 & 14 & 30\\
1742$-$3957 & 17:42:04.43(19) & $-$39:57:22(9) & 350.21 & $-5.13$ & 7 & 1.03 & 18.0 & 0.14 & 65 & 120\\
1744$-$3922 & 17:44:02.675(10) & $-$39:22:21.1(4) & 350.91 & $-5.15$ & 4 & 1.26 & 26.5 & 0.20 & 3.4 & 7.3\\
1750$-$2043 & 17:50:18.4(5) & $-$20:43:08(75) & 7.59 & $3.33$ & 10 & 0.58 & 21.1 & 0.24 & 358 & 432\\
1751$-$2857 & 17:51:32.6965(2) & $-$28:57:46.50(3) & 0.65 & $-1.12$ & 11 & 0.91 & 18.5 & 0.06 & 0.25 & 0.74\\
\\
1753$-$1914 & 17:53:35.171(12) & $-$19:14:58(3) & 9.25 & $3.41$ & 2 & 0.23 & 49.4 & 0.13 & 10 & 14\\
1754$-$3510 & 17:54:54.616(7) & $-$35:10:43.0(5) & 355.64 & $-4.88$ & 4 & 0.50 & 101.8 & 0.47 & 4.1 & 13\\
1755$-$1650 & 17:55:11.648(18) & $-$16:50:41(3) & 11.53 & $4.29$ & 4 & 0.52 & 17.0 & 0.13 & 14 & 26\\
1756$-$2251 & 17:56:46.6332(2) & $-$22:51:59.4(2) & 6.50 & $0.95$ & 7 & 0.87 & 19.6 & 0.6 & 0.78 & 1.6\\
1758$-$2846 & 17:58:15.41(3) & $-$28:46:02(5) & 1.56 & $-2.29$ & 7 & 0.65 & 32.4 & 0.20 & 4.4 & 30\\
\\
1759$-$1736 & 17:59:28.16(7) & $-$17:36:10(7) & 11.38 & $3.03$ & 9 & 0.26 & 11.2 & 0.12 & 18 & 44\\
1801$-$1417 & 18:01:51.0764(5) & $-$14:17:34.50(5) & 14.55 & $4.16$ & 12 & 1.28 & 37.7 & 0.17 & 0.6 & 1.3\\
1801$-$3458 & 18:01:52.66(11) & $-$34:58:37(7) & 356.52 & $-6.02$ & 12 & 0.65 & 12.7 & 0.11 & 47 & 110\\
1802$-$2124 & 18:02:05.3352(3) & $-$21:24:03.6(3) & 8.38 & $0.61$ & 4 & 0.93 & 15.2 & 0.77 & 0.37 & 0.74\\
1805$-$1504 & 18:05:06.10(17) & $-$15:04:36(18) & 14.25 & $3.09$ & 7 & 1.10 & 94.8 & 2.2 & 229 & 340\\
\\
1808$-$1020 & 18:08:45.634(6) & $-$10:20:48.3(4) & 18.83 & $4.60$ & 9 & 0.89 & 34.9 & 0.23 & 8 & 23\\
1808$-$1726 & 18:08:42.3(3) & $-$17:26:27(42) & 12.60 & $1.19$ & 5 & 0.28 & 18.9 & 0.39 & 66 & --\\
1808$-$2701 & 18:08:13.23(4) & $-$27:01:21(9) & 4.16 & $-3.36$ & 11 & 0.88 & 16.5 & 0.15 & 23 & 80\\
1811$-$2439 & 18:11:55.53(3) & $-$24:39:53(9) & 6.63 & $-2.95$ & 1 & 0.66 & 37.3 & 0.26 & 15 & 24\\
1812$-$2526 & 18:12:32.30(3) & $-$25:26:38(7) & 6.01 & $-3.45$ & 11 & 0.80 & 13.1 & 0.18 & 12 & 19\\
\\
1814$-$0618 & 18:14:41.26(7) & $-$06:18:01.7(33) & 23.10 & $5.23$ & 8 & 0.45 & 16.9 & 0.58 & 48 & 194\\
1816$-$0755 & 18:16:24.543(6) & $-$07:55:22.5(4) & 21.87 & $4.09$ & 8 & 0.85 & 23.7 & 0.17 & 6 & 17\\
1817$-$0743 & 18:17:49.79(5) & $-$07:43:18.9(14) & 22.21 & $3.88$ & 9 & 0.32 & 13.6 & 0.25 & 13 & 34\\
1819$-$1318 & 18:19:43.66(5) & $-$13:18:42(5) & 17.50 & $0.83$ & 13 & 0.46 & 13.7 & 0.16 & 23 & 41\\
1820$-$0509 & 18:20:22.728(7) & $-$05:09:38.5(4) & 24.78 & $4.52$ & 5 & 0.63 & 32.4 & 0.25 & 9 & 18\\
\\
1821$-$0256 & 18:21:10.310(20) & $-$02:56:38.6(10) & 26.84 & $5.37$ & 1 & 0.98 & 9.7 & 0.19 & 10 & 25\\
1822$-$0848 & 18:22:51.86(8) & $-$08:48:59.0(11) & 21.83 & $2.26$ & 1 & 0.35 & 28.7 & 0.04 & 18 & 55\\
1824$-$0127 & 18:24:53.43(4) & $-$01:27:51.4(18) & 28.59 & $5.23$ & 4 & 1.18 & 31.7 & 0.59 & 20 & 81\\
1824$-$2233 & 18:24:10.32(5) & $-$22:33:11(16) & 9.82 & $-4.44$ & 7 & 0.34 & 49.6 & 0.22 & 13 & 32\\
1824$-$2328 & 18:24:28.64(19) & $-$23:28:17(61) & 9.04 & $-4.92$ & 2 & 1.12 & 61.0 & 0.32 & 35 & 51\\
\\
1827$-$0750 & 18:27:02.735(6) & $-$07:50:19.2(6) & 23.18 & $1.80$ & 1 & 0.38 & 119.1 & 1.4 & 10 & 36\\
1828$-$2119 & 18:28:21.56(5) & $-$21:19:56(11) & 11.36 & $-4.73$ & 4 & 0.75 & 37.5 & 0.38 & 22 & 34\\
1829$+$0000 & 18:29:47.14(3) & $+$00:00:08.5(11) & 30.46 & $4.82$ & 6 & 1.07 & 20.4 & 0.43 & 16 & 30\\
1830$-$0052 & 18:30:30.70(5) & $-$00:52:54.0(15) & 29.76 & $4.25$ & 4 & 0.23 & 10.5 & 0.04 & 9 & 16\\
1830$-$0131 & 18:30:19.612(12) & $-$01:31:48.1(6) & 29.16 & $3.99$ & 7 & 1.43 & 13.1 & 0.35 & 11 & 18\\
\\
1830$-$1414 & 18:30:00.35(4) & $-$14:14:37(4) & 17.84 & $-1.81$ & 2 & 0.18 & 11.6 & 0.10 & 10 & 22\\
1831$-$0952 & 18:31:34.304(16) & $-$09:52:01.7(11) & 21.90 & $-0.13$ & 11 & 0.59 & 13.9 & 0.33 & 8 & 14\\
1832$+$0029 & 18:32:50.7(3) & $+$00:29:27(18) & 31.25 & $4.36$ & 10 & 0.87 & 12.5 & 0.14 & 7 & 15\\
1834$-$0031 & 18:34:51.093(14) & $-$00:31:34.3(6) & 30.57 & $3.45$ & 4 & 1.04 & 14.3 & 0.17 & 5 & 13\\
1834$-$1855 & 18:34:45.94(5) & $-$18:55:59(6) & 14.19 & $-4.98$ & 5 & 1.00 & 27.2 & 0.48 & 32 & 71\\
\\
1835$-$0349 & 18:35:12.946(12) & $-$03:49:09.9(9) & 27.68 & $1.86$ & 7 & 0.57 & 26.6 & 0.16 & 15 & 28\\
1835$-$0944 & 18:35:46.66(9) & $-$09:44:29(3) & 22.49 & $-0.99$ & 1 & 0.30 & 18.8 & 0.41 & 7 & 19\\
1835$-$0946 & 18:35:44.86(9) & $-$09:46:38(3) & 22.45 & $-1.00$ & 1 & 0.56 & 21.2 & 0.18 & 4 & 7.2\\
1835$-$1548 & 18:35:50.33(10) & $-$15:48:38(10) & 17.09 & $-3.78$ & 6 & 0.51 & 9.7 & 0.06 & 17 & 42\\
1836$-$1324 & 18:36:52.287(9) & $-$13:24:33.6(5) & 19.35 & $-2.91$ & 4 & 3.83 & 20.0 & 0.10 & 2.4 & 41\\
\\

\hline 
\end{tabular}\end{footnotesize}\end{center}\end{table*}
\addtocounter{table}{-1}\begin{table*} 
\begin{center}\begin{footnotesize}
\caption{-- {\it continued}}
\begin{tabular}{lllrrccrlrr}
\hline
PSR J & R.A. (J2000) & Dec. (J2000) & \multicolumn{1}{c}{$l$} & \multicolumn{1}{c}{$b$} & Beam & Radial & S/N & $S_{1400}$ & W$_{50}$ & W$_{10}$ \\
 & (h~~~m~~~s) & (\degr ~~~\arcmin ~~~\arcsec) & \multicolumn{1}{c}{(\degr)} &\multicolumn{1}{c}{(\degr)} & & Distance & & (mJy) & (ms) & (ms) \\ 
\hline 
1837$+$0053 & 18:37:28.45(7) & $+$00:53:13(3) & 32.13 & $3.51$ & 7 & 1.17 & 18.7 & 0.34 & 49 & 108\\
1839$-$0436 & 18:39:00.064(9) & $-$04:36:57.5(7) & 27.41 & $0.65$ & 8 & 0.65 & 10.9 & 0.23 & 6 & 12\\
1839$-$1238 & 18:39:43.708(19) & $-$12:38:40.7(18) & 20.35 & $-3.18$ & 4 & 3.73 & 75.7 & 0.37 & 30 & 53\\
1840$+$0214 & 18:40:34.087(19) & $+$02:14:54.7(6) & 33.70 & $3.44$ & 3 & 0.31 & 13.8 & 0.07 & 10 & 22\\
1840$-$0840 & 18:40:51.9(4) & $-$08:40:29(15) & 24.01 & $-1.62$ & 2 & 0.03 & 61.7 & 1.00 & 188 & 343\\
\\
1840$-$1207 & 18:40:53.782(8) & $-$12:07:32.6(6) & 20.94 & $-3.20$ & 8 & 0.77 & 24.2 & 0.22 & 15 & 22\\
1841$+$0130 & 18:41:17.761(8) & $+$01:30:16.9(3) & 33.12 & $2.94$ & 12 & 1.06 & 11.4 & 0.06 & 6 & 11\\
1841$-$1404 & 18:41:34.3(4) & $-$14:04:19(41) & 19.28 & $-4.23$ & 7 & 0.15 & 19.0 & 0.18 & 17 & 122\\
1842$+$0257 & 18:42:30.93(4) & $+$02:57:58.8(14) & 34.56 & $3.34$ & 4 & 1.06 & 55.0 & 0.26 & 52 & 85\\
1842$+$0358 & 18:42:17.022(6) & $+$03:58:35.3(3) & 35.43 & $3.85$ & 6 & 0.63 & 20.2 & 0.09 & 4 & 8\\
\\
1842$+$0638 & 18:42:10.002(17) & $+$06:38:15.0(7) & 37.80 & $5.07$ & 10 & 0.98 & 16.1 & 0.21 & 13 & 20\\
1843$-$1507 & 18:43:33.607(7) & $-$15:07:03.0(8) & 18.56 & $-5.13$ & 9 & 0.63 & 46.1 & 0.17 & 7 & 18\\
1845$+$0623 & 18:45:08.59(3) & $+$06:23:57.6(9) & 37.92 & $4.31$ & 1 & 0.53 & 39.0 & 0.33 & 13 & --\\
1845$-$0826 & 18:45:05.693(19) & $-$08:26:39.7(12) & 24.69 & $-2.44$ & 10 & 0.57 & 15.4 & 0.33 & 13 & 31\\
1845$-$1114 & 18:45:45.778(6) & $-$11:14:11.0(5) & 22.28 & $-3.86$ & 2 & 0.27 & 41.9 & 0.52 & 4 & 7\\
\\
1845$-$1351 & 18:45:11.50(3) & $-$13:51:54.5(20) & 19.86 & $-4.92$ & 2 & 0.61 & 26.2 & 0.33 & 63 & 84\\
1846$-$0749 & 18:46:19.744(5) & $-$07:49:21.4(3) & 25.39 & $-2.43$ & 12 & 1.49 & 15.9 & 0.35 & 7 & 15\\
1848$-$0601 & 18:48:20.349(14) & $-$06:01:07.7(8) & 27.22 & $-2.05$ & 4 & 0.74 & 18.3 & 0.24 & 9 & 22\\
1848$-$1150 & 18:48:11.81(3) & $-$11:50:09.2(18) & 22.01 & $-4.66$ & 10 & 0.40 & 34.6 & 0.21 & 31 & 42\\
1849$+$0409 & 18:49:03.471(13) & $+$04:09:42.3(5) & 36.37 & $2.42$ & 4 & 0.23 & 33.1 & 0.10 & 9 & 17\\
\\
1851$-$0114 & 18:51:16.41(3) & $-$01:14:13.4(13) & 31.81 & $-0.53$ & 2 & 0.87 & 18.5 & 0.28 & 22 & 44\\
1853$+$0853 & 18:53:22.07(11) & $+$08:53:17(3) & 41.07 & $3.61$ & 5 & 0.35 & 19.6 & 0.11 & 50 & 90\\
1853$+$1303 & 18:53:57.31827(8) & $+$13:03:44.0884(17) & 44.87 & $5.37$ & 1 & 0.85 & 15.1 & 0.4 & 0.59 & 2.19\\
1856$-$0526 & 18:56:21.977(15) & $-$05:26:56.8(8) & 28.64 & $-3.57$ & 13 & 0.86 & 35.1 & 0.40 & 28 & 39\\
1901$-$0315 & 19:01:16.33(8) & $-$03:15:14.3(20) & 31.15 & $-3.67$ & 2 & 0.96 & 11.2 & 0.09 & 10 & 20\\
\\
1901$+$0435 & 19:01:32.2(3) & $+$04:35:23(10) & 38.17 & $-0.15$ & 5 & 0.43 & 43.8 & 0.27 & 315 & --\\
1901$+$0621 & 19:01:05.96(9) & $+$06:21:19(4) & 39.69 & $0.76$ & 10 & 0.76 & 14.2 & 0.47 & 43 & --\\
1903$-$0258 & 19:03:30.343(11) & $-$02:58:15.6(7) & 31.66 & $-4.04$ & 9 & 0.37 & 22.3 & 0.14 & 7 & --\\
1903$+$0925 & 19:03:55.18(20) & $+$09:25:55(6) & 42.74 & $1.54$ & 5 & 1.19 & 21.7 & 0.20 & 89 & --\\
1904$-$0150 & 19:04:55.555(10) & $-$01:50:32.5(5) & 32.83 & $-3.84$ & 7 & 0.33 & 23.0 & 0.09 & 5 & 13\\
\\
1906$+$0414 & 19:06:57.793(7) & $+$04:14:29.44(13) & 38.48 & $-1.51$ & 6 & 0.79 & 12.7 & 0.23 & 20 & --\\
1907$+$1149 & 19:07:37.78(6) & $+$11:49:37.1(17) & 45.29 & $1.83$ & 11 & 0.55 & 25.3 & 0.13 & 15 & 30\\
1910$-$0112 & 19:10:15.7(4) & $-$01:12:06(14) & 34.01 & $-4.74$ & 10 & 0.66 & 11.0 & 0.07 & 50 & 89\\
1910$+$1256 & 19:10:09.70041(6) & $+$12:56:25.5276(6) & 46.56 & $1.80$ & 12 & 0.16 & 17.2 & 0.5 & 0.14 & 0.43\\
1911$+$1347 & 19:11:55.2047(3) & $+$13:47:34.411(7) & 47.52 & $1.81$ & 1 & 0.31 & 36.3 & 0.08 & 0.15 & 0.44\\
\\
1913$+$0904 & 19:13:21.061(8) & $+$09:04:45.4(4) & 43.50 & $-0.68$ & 4 & 0.54 & 33.3 & 0.07 & 3 & 6\\
1914$+$0219 & 19:14:23.794(8) & $+$02:19:47.8(3) & 37.63 & $-4.04$ & 13 & 0.62 & 172.0 & 0.75 & 16 & 24\\
1915$+$0227 & 19:15:02.122(5) & $+$02:27:47.77(16) & 37.82 & $-4.12$ & 9 & 0.69 & 56.3 & 0.55 & 7 & 27\\
1915$+$1410 & 19:15:35.416(12) & $+$14:10:51.5(3) & 48.27 & $1.20$ & 9 & 0.47 & 17.7 & 0.10 & 15 & 24\\
1921$+$0812 & 19:21:47.704(5) & $+$08:12:51.86(10) & 43.71 & $-2.93$ & 11 & 0.64 & 30.0 & 0.66 & 2 & 5\\
\\
1927$+$0911 & 19:27:25.628(8) & $+$09:11:05.7(3) & 45.22 & $-3.70$ & 2 & 0.78 & 14.1 & 0.16 & 3 & 8\\
1941$+$1341 & 19:41:04.867(18) & $+$13:41:30.4(6) & 50.80 & $-4.47$ & 9 & 0.57 & 17.7 & 0.17 & 9 & 33\\

\hline 
\end{tabular}\end{footnotesize}\end{center}\end{table*}

\begin{table*} 
\caption{Periods, period derivatives and dispersion measures for 142
pulsars discovered in this phase of the Parkes multibeam pulsar
survey. We also give the MJD of the epoch used for period
determination, the number of TOAs included in the timing solution, the
MJD range covered and the RMS of the post-fit timing residuals.}
\label{tb:prd}
\begin{center}\begin{footnotesize}
\begin{tabular}{llllllll}
\hline 
PSR J & Period, $P$ & $\dot{P}$ & Epoch & N$_{\rm TOA}$ & Data Span & Residual & DM \\ 
      & (s)       & (10$^{-15}$) & (MJD) &         & (MJD)     & (ms)     & (cm$^{-3}$pc) \\ 
\hline 
0733$-$2345 & 1.79624970117(10) & 5.704(10) & 53014.0 & 27 & 52574-53454 & 1.8 & 194(3)\\ 
0932$-$5327 & 4.3921587127(6) & 8.37(8) & 53099.0 & 21 & 52772-53426 & 4.0 & 122(11)\\ 
1001$-$5939 & 7.7336402649(8) & 59.9(3) & 53069.0 & 39 & 52826-53312 & 2.0 & 113(3)\\ 
1020$-$6026 & 0.140479910179(8) & 6.7386(20) & 53128.0 & 31 & 52854-53403 & 1.4 & 445.0(20)\\ 
1107$-$5907 & 0.252773323418(6) & 0.009(10) & 53089.0 & 53 & 52698-53480 & 0.8 & 40.2(11)\\ 
\\
1125$-$6014 & 0.0026303807397848(8) & 0.00000401(9) & 53141.0 & 36 & 52827-53454 & 0.006 & 52.951(14)\\ 
1148$-$5725 & 3.5599372320(7) & 11.1(13) & 52837.0 & 20 & 52569-53104 & 2.6 & 174(5)\\ 
1216$-$6410 & 0.003539375658423(3) & 0.00000162(19) & 53391.0 & 40 & 53051-53732 & 0.011 & 47.40(3)\\ 
1308$-$5844 & 0.4646997488(7) & 8.639(5) & 51560.0 & 24 & 52692-53104 & 0.4 & 205.6(11)\\ 
1355$-$5747 & 2.0386738254(4) & 0.71(9) & 53513.0 & 31 & 53283-53742 & 3.3 & 229(3)\\ 
\\
1357$-$6429 & 0.16610832750(1) & 360.184(1) & 52921.0 & 28 & 52738-53104 & 0.9 & 127.2(5)\\ 
1405$-$5641 & 0.61757468514(4) & 1.198(12) & 52898.0 & 21 & 52692-53104 & 0.6 & 273(3)\\ 
1439$-$5501 & 0.028634888190455(12) & 0.0001418(10) & 53200.0 & 170 & 52915-53716 & 0.025 & 14.56(5)\\ 
1519$-$5734 & 0.51875778474(3) & 4.273(5) & 53105.0 & 32 & 52807-53403 & 1.8 & 664(3)\\ 
1519$-$6106 & 2.15430706064(14) & 8.372(19) & 53048.0 & 36 & 52695-53402 & 1.8 & 221(3)\\ 
\\
1538$-$5732 & 0.341212811806(6) & 4.5498(6) & 51803.0 & 40 & 51300-52306 & 1.0 & 152.7(8)\\ 
1558$-$5756 & 1.12234168628(3) & 186.406(4) & 53139.0 & 41 & 52855-53422 & 0.9 & 127.8(10)\\ 
1600$-$5916 & 1.2476661740(3) & 0.89(7) & 53116.0 & 31 & 52807-53425 & 6.7 & 177(4)\\ 
1604$-$4718 & 0.527465874211(20) & 4.961(9) & 52897.0 & 20 & 52695-53100 & 0.5 & 52.0(16)\\ 
1611$-$5847 & 0.3545503166464(18) & 0.002(4) & 53071.0 & 33 & 52739-53403 & 0.2 & 79.9(13)\\ 
\\
1617$-$4608 & 0.567080221721(10) & 16.4665(10) & 53129.0 & 34 & 52855-53403 & 0.6 & 321.5(12)\\ 
1622$-$4347 & 0.457681471040(9) & 5.051(4) & 52897.0 & 19 & 52695-53100 & 0.2 & 177(3)\\ 
1624$-$4411 & 0.233164076390(4) & 0.7992(8) & 53058.0 & 38 & 52695-53422 & 0.6 & 139.4(14)\\ 
1624$-$4613 & 0.8712426056(4) & $<$0.12 & 53500.0 & 20 & 53284-53716 & 4.2 & 224.2(19)\\ 
1627$-$4706 & 0.140745806053(6) & 1.7305(6) & 53165.0 & 62 & 52807-53523 & 1.7 & 456.1(18)\\ 
\\
1631$-$4155 & 0.55124125377(4) & 0.0088(73) & 53082.0 & 38 & 52739-53425 & 1.8 & 235(4)\\ 
1634$-$5107 & 0.50735578252(6) & 1.573(3) & 52860.0 & 49 & 52294-53426 & 3.4 & 372.8(20)\\ 
1637$-$4450 & 0.252869846841(20) & 0.5755(20) & 53283.0 & 28 & 53078-53489 & 0.4 & 470.7(8)\\ 
1638$-$3815 & 0.69826060054(20) & 0.077(3) & 52135.0 & 22 & 52601-53107 & 0.7 & 238.0(11)\\ 
1638$-$4725 & 0.7639335(3) & 4.8(9) & 52651.0 & 109 & 51856-53445 & 5.3 & 552.1(14)\\ 
\\
1643$-$4505 & 0.237383081875(17) & 31.833(10) & 52923.0 & 17 & 52739-53107 & 0.9 & 484(3)\\ 
1646$-$5123 & 0.530075207410(20) & 2.099(12) & 52906.0 & 17 & 52739-53100 & 1.0 & 279(3)\\ 
1649$-$3805 & 0.262025692969(7) & 0.0367(14) & 52885.0 & 21 & 52695-53100 & 0.3 & 213.8(5)\\ 
1650$-$4921 & 0.1563993674972(10) & 1.81873(20) & 53253.0 & 45 & 52983-53523 & 0.2 & 229.9(5)\\ 
1653$-$4030 & 1.01937155664(16) & 0.44(4) & 53139.0 & 36 & 52855-53422 & 5.8 & 425(8)\\ 
\\
1654$-$3710 & 0.93916547876(3) & 0.735(3) & 53131.0 & 36 & 52859-53403 & 1.3 & 302.0(12)\\ 
1656$-$3621 & 0.730133707454(16) & 1.2752(16) & 53123.0 & 39 & 52824-53422 & 1.4 & 229(3)\\ 
1700$-$3611 & 1.49409061528(16) & 4.324(16) & 52828.0 & 27 & 52584-53100 & 1.2 & 232.7(12)\\ 
1700$-$4012 & 0.283791853785(11) & 0.099(4) & 53105.0 & 35 & 52808-53403 & 1.7 & 385(4)\\ 
1711$-$4322 & 0.1026182883472(6) & 0.02666(5) & 52433.0 & 98 & 51412-53454 & 0.8 & 191.5(7)\\ 
\\
1715$-$3859 & 0.92810750775(3) & 4.397(7) & 53097.0 & 35 & 52772-53422 & 6.4 & 817(5)\\ 
1717$-$3953 & 1.08552061541(18) & 0.033(19) & 53313.0 & 38 & 52884-53742 & 9.6 & 466(8)\\ 
1718$-$4539 & 0.59047278772(3) & 7.507(7) & 53105.0 & 29 & 52808-53403 & 1.0 & 254(6)\\ 
1722$-$4400 & 0.2185540958272(15) & 0.3738(4) & 53058.0 & 36 & 52695-53422 & 0.3 & 219.3(5)\\ 
1724$-$4500 & 1.3091087873(3) & 3.51(12) & 53255.0 & 24 & 53052-53458 & 3.3 & 182(3)\\ 
\\
1725$-$2852 & 1.257787625(10) & 1.99(15) & 52197.0 & 20 & 52692-53103 & 3.7 & 161(10)\\ 
1728$-$4028 & 0.866342509(4) & 0.23(6) & 52077.0 & 20 & 52692-53103 & 2.4 & 231(10)\\ 
1730$-$2900 & 1.5384267263(3) & 8.85(8) & 53128.0 & 26 & 52854-53402 & 2.9 & 289(5)\\ 
1732$-$3426 & 0.33288667881(4) & 0.402(9) & 52359.0 & 23 & 52146-52572 & 1.2 & 513.5(18)\\ 
1733$-$2533 & 0.65979414168(4) & 2.1(5) & 53070.0 & 32 & 52739-53402 & 2.5 & 242(5)\\ 
\\

\hline 
\end{tabular}\end{footnotesize}\end{center}\end{table*}
\addtocounter{table}{-1}\begin{table*} 
\begin{center}\begin{footnotesize}
\caption{-- {\it continued}}\begin{tabular}{llllllll}
\hline 
PSR J & Period, $P$ & $\dot{P}$ & Epoch & N$_{\rm TOA}$ & Data Span & Residual & DM \\ 
      & (s)       & (10$^{-15}$) & (MJD) &         & (MJD)     & (ms)     & (cm$^{-3}$pc) \\ 
\hline 
1733$-$2837 & 0.76818455674(6) & 1.108(8) & 53128.0 & 22 & 52854-53402 & 1.7 & 225(5)\\ 
1734$-$2415 & 0.61252373999(12) & 1.051(12) & 53164.0 & 56 & 52856-53471 & 6.4 & 126.3(7)\\ 
1736$-$2457 & 2.64222343904(14) & 3.43(3) & 53095.0 & 50 & 52718-53471 & 3.5 & 170(4)\\ 
1738$-$2330 & 1.97884743163(4) & 8.559(4) & 53206.0 & 33 & 52695-53716 & 1.0 & 99.3(18)\\ 
1739$-$3951 & 0.341772476799(10) & 0.02(5) & 52897.0 & 17 & 52692-53103 & 0.4 & 24.6(4)\\ 
\\
1741$-$2719 & 0.346796929142(19) & 0.081(4) & 52836.0 & 23 & 52569-53103 & 1.2 & 361.9(4)\\ 
1742$-$3957 & 1.0163491951(4) & 0.082(29) & 53124.0 & 29 & 52825-53422 & 7.9 & 186(8)\\ 
1744$-$3922 & 0.172444360995(2) & 0.00155(12) & 52530.0 & 36 & 51953-53107 & 0.3 & 148.1(7)\\ 
1750$-$2043 & 5.639047079(3) & 7.9(16) & 52923.0 & 18 & 52739-53107 & 6.2 & 239(7)\\ 
1751$-$2857 & 0.0039148731963690(6) & 0.00001126(4) & 52560.0 & 168 & 51808-53312 & 0.029 & 42.808(20)\\ 
\\
1753$-$1914 & 0.0629548889725(12) & 0.00202(12) & 53150.0 & 62 & 52828-53471 & 0.8 & 105.3(3)\\ 
1754$-$3510 & 0.392703893399(3) & 0.7816(5) & 53097.0 & 43 & 52733-53461 & 0.4 & 82.3(3)\\ 
1755$-$1650 & 0.73374443871(3) & 0.686(3) & 53165.0 & 55 & 52860-53471 & 1.1 & 159.9(10)\\ 
1756$-$2251 & 0.02846158845494(2) & 0.0010171(2) & 52086.0 & 1597 & 50997-53176 & 0.042 & 121.18(2)\\ 
1758$-$2846 & 0.76670638621(3) & 0.094(18) & 53253.0 & 39 & 53063-53443 & 1.3 & 66.6(3)\\ 
\\
1759$-$1736 & 0.79845156695(8) & 0.255(20) & 53156.0 & 18 & 52854-53458 & 1.7 & 206(3)\\ 
1801$-$1417 & 0.0036250966601209(18) & 0.00000531(11) & 53156.0 & 89 & 52569-53743 & 0.021 & 57.21(4)\\ 
1801$-$3458 & 1.3856036231(3) & 0.4(4) & 53112.0 & 27 & 52827-53397 & 4.5 & 146(6)\\ 
1802$-$2124 & 0.012647593582763(5) & 0.000072(1) & 52855.0 & 30 & 52605-53105 & 0.007 & 149.6(1)\\ 
1805$-$1504 & 1.1812692298(3) & 0.27(3) & 53035.0 & 58 & 52599-53471 & 14.2 & 225(3)\\ 
\\
1808$-$1020 & 0.596993202504(4) & 0.7725(4) & 53035.0 & 63 & 52599-53471 & 0.5 & 225.3(8)\\ 
1808$-$1726 & 0.24103454795(9) & $<$0.012 & 53108.0 & 67 & 52744-53471 & 23.2 & 536(7)\\ 
1808$-$2701 & 2.45788177758(12) & 65.818(11) & 53112.0 & 56 & 52751-53472 & 2.6 & 95(4)\\ 
1811$-$2439 & 0.415812941402(11) & 0.2972(11) & 53165.0 & 54 & 52859-53471 & 0.9 & 172.0(5)\\ 
1812$-$2526 & 0.315835048197(8) & 0.1775(9) & 53171.0 & 54 & 52871-53471 & 1.1 & 361.4(4)\\ 
\\
1814$-$0618 & 1.37786849563(14) & 0.292(16) & 53095.0 & 66 & 52718-53471 & 5.7 & 168(6)\\ 
1816$-$0755 & 0.2176426911444(15) & 6.48032(15) & 53035.0 & 60 & 52599-53471 & 0.5 & 116.8(4)\\ 
1817$-$0743 & 0.438095346909(18) & $<$0.0083 & 52887.0 & 13 & 52667-53107 & 0.7 & 14.8(4)\\ 
1819$-$1318 & 1.51569599263(12) & 0.6(3) & 52856.0 & 13 & 52604-53107 & 1.0 & 35.1(15)\\ 
1820$-$0509 & 0.337320795791(3) & 0.9323(3) & 53035.0 & 47 & 52608-53461 & 0.4 & 104.0(3)\\ 
\\
1821$-$0256 & 0.414111049843(14) & 0.0372(14) & 53097.0 & 67 & 52732-53461 & 1.8 & 84.0(4)\\ 
1822$-$0848 & 0.834839272621(14) & 0.135(4) & 53115.0 & 229 & 52749-53481 & 1.5 & 186.3(7)\\ 
1824$-$0127 & 2.49946826743(11) & 3.911(11) & 53040.0 & 51 & 52608-53471 & 3.0 & 58.0(15)\\ 
1824$-$2233 & 1.16174310971(3) & 0.301(3) & 53165.0 & 52 & 52859-53471 & 0.8 & 156.5(12)\\ 
1824$-$2328 & 1.50587455169(3) & 1.74(3) & 53111.0 & 44 & 52751-53471 & 0.7 & 185(3)\\ 
\\
1827$-$0750 & 0.270502039371(3) & 1.5446(8) & 53223.0 & 38 & 52974-53471 & 0.4 & 381(9)\\ 
1828$-$2119 & 0.514523069504(14) & 1.2668(14) & 53035.0 & 57 & 52599-53471 & 1.9 & 268.0(6)\\ 
1829$+$0000 & 0.199147397053(8) & 0.5249(8) & 53112.0 & 75 & 52752-53471 & 2.5 & 114.0(4)\\ 
1830$-$0052 & 0.34569803402(5) & 0.237(8) & 53276.0 & 17 & 53063-53489 & 0.8 & 220.4(9)\\ 
1830$-$0131 & 0.152511958009(4) & 2.106(4) & 53095.0 & 125 & 52718-53471 & 1.5 & 95.7(3)\\ 
\\
1830$-$1414 & 0.77149252038(6) & 0.076(11) & 52265.0 & 19 & 52003-52528 & 1.5 & 393.6(14)\\ 
1831$-$0952 & 0.0672668461152(14) & 8.32385(3) & 52412.0 & 124 & 51302-53523 & 2.0 & 247(5)\\ 
1832$+$0029 & 0.533917296(6) & 1.51(20) & 53344.0 & 13 & 52887-53153 & 0.7 & 28.3(12)\\ 
1834$-$0031 & 0.329531922488(6) & 0.4486(6) & 53030.0 & 48 & 52599-53461 & 1.0 & 155.1(6)\\ 
1834$-$1855 & 1.46565577133(7) & 1.931(9) & 53108.0 & 40 & 52744-53471 & 1.8 & 185.2(12)\\ 
\\
1835$-$0349 & 0.84186451859(4) & 3.058(4) & 53125.0 & 20 & 52853-53397 & 0.5 & 269.6(7)\\ 
1835$-$0944 & 0.145346822521(15) & 4.39(13) & 53663.9 & 14 & 53525-53802 & 0.5 & 277.2(5)\\ 
1835$-$0946 & 0.37953610601(5) & 0.043(31) & 53663.0 & 13 & 53525-53802 & 0.4 & 193.3(5)\\ 
1835$-$1548 & 0.67048566238(9) & 1.74(14) & 53140.0 & 21 & 52856-53425 & 3.1 & 327(6)\\ 
1836$-$1324 & 0.1787563517731(16) & 1.0366(10) & 53277.0 & 41 & 53083-53471 & 0.2 & 157.33(10)\\ 
\\

\hline 
\end{tabular}\end{footnotesize}\end{center}\end{table*}
\addtocounter{table}{-1}\begin{table*} 
\begin{center}\begin{footnotesize}
\caption{-- {\it continued}}\begin{tabular}{llllllll}
\hline 
PSR J & Period, $P$ & $\dot{P}$ & Epoch & N$_{\rm TOA}$ & Data Span & Residual & DM \\ 
      & (s)       & (10$^{-15}$) & (MJD) &         & (MJD)     & (ms)     & (cm$^{-3}$pc) \\ 
\hline 
1837$+$0053 & 0.47351252978(4) & 0.038(4) & 53094.0 & 25 & 52694-53494 & 2.8 & 124(3)\\ 
1839$-$0436 & 0.149460655762(3) & 0.8096(3) & 53190.0 & 29 & 52887-53493 & 0.5 & 292.7(18)\\ 
1839$-$1238 & 1.91142802251(5) & 4.947(4) & 53035.0 & 46 & 52599-53471 & 1.4 & 169.8(18)\\ 
1840$+$0214 & 0.79747807530(4) & 8.294(7) & 53129.0 & 19 & 52860-53397 & 0.5 & 182.4(10)\\ 
1840$-$0840 & 5.3093766847(20) & 23.7(12) & 53278.0 & 39 & 53084-53471 & 9.3 & 272(19)\\ 
\\
1840$-$1207 & 0.754470427192(6) & 3.1972(6) & 53038.0 & 57 & 52604-53471 & 0.5 & 302.3(15)\\ 
1841$+$0130 & 0.0297727753332(4) & 0.00817(5) & 53111.0 & 155 & 52749-53472 & 0.7 & 125.88(6)\\ 
1841$-$1404 & 1.3345572305(8) & 0.64(18) & 53173.0 & 21 & 52856-53491 & 5.9 & 267(8)\\ 
1842$+$0257 & 3.08825579295(13) & 29.591(12) & 53040.0 & 44 & 52608-53471 & 2.3 & 148.1(11)\\ 
1842$+$0358 & 0.233326206679(3) & 0.8115(3) & 53108.0 & 56 & 52744-53471 & 0.5 & 109.9(4)\\ 
\\
1842$+$0638 & 0.313016401038(11) & 0.076(17) & 53125.0 & 18 & 52854-53397 & 0.4 & 212.2(12)\\ 
1843$-$1507 & 0.583550377362(5) & 7.1999(6) & 53095.0 & 49 & 52718-53471 & 0.5 & 215.5(7)\\ 
1845$+$0623 & 1.42165378409(7) & 0.546(7) & 53095.0 & 42 & 52718-53471 & 1.8 & 113.0(14)\\ 
1845$-$0826 & 0.634354346767(17) & 9.353(3) & 53112.0 & 40 & 52752-53471 & 1.3 & 228.2(12)\\ 
1845$-$1114 & 0.2062199811321(13) & 2.00552(13) & 53036.0 & 58 & 52599-53472 & 0.4 & 206.7(5)\\ 
\\
1845$-$1351 & 2.61891846726(6) & 9.723(8) & 53112.0 & 39 & 52751-53472 & 1.2 & 197.4(14)\\ 
1846$-$0749 & 0.350109572852(4) & 1.2616(4) & 53172.0 & 51 & 52871-53472 & 0.3 & 388.3(5)\\ 
1848$-$0601 & 0.225004475404(4) & 0.2871(4) & 53041.0 & 48 & 52609-53472 & 1.0 & 496.6(4)\\ 
1848$-$1150 & 1.31221843801(3) & 1.434(3) & 53040.0 & 51 & 52608-53472 & 1.5 & 163.4(18)\\ 
1849$+$0409 & 0.761194081655(11) & 21.5864(12) & 53035.0 & 41 & 52608-53461 & 0.8 & 56.1(14)\\ 
\\
1851$-$0114 & 0.95318168518(4) & 2.483(4) & 53038.0 & 48 & 52604-53472 & 2.2 & 427.2(14)\\ 
1853$+$0853 & 3.9146579157(5) & 5.13(7) & 53097.0 & 17 & 52739-53455 & 2.9 & 214(5)\\ 
1853$+$1303 & 0.0040917973806819(14) & 0.00000885(10) & 52972.0 & 140 & 52606-53337 & 0.003 & 30.5702(12)\\ 
1856$-$0526 & 0.370483417429(7) & 1.6975(9) & 53095.0 & 42 & 52718-53472 & 1.0 & 130.5(4)\\ 
1901$-$0315 & 0.80169309688(11) & 2.57(4) & 53283.0 & 22 & 53075-53491 & 1.0 & 242.6(16)\\ 
\\
1901$+$0435 & 0.6905763581(3) & 8.67(3) & 53041.0 & 118 & 52600-53481 & 33.8 & 1042.6(10)\\ 
1901$+$0621 & 0.83200194892(5) & 0.018(3) & 52282.0 & 21 & 51458-53106 & 4.2 & 94(7)\\ 
1903$-$0258 & 0.301458774079(6) & 0.6791(6) & 53095.0 & 49 & 52718-53472 & 0.9 & 113.0(5)\\ 
1903$+$0925 & 0.35715482026(9) & 36.899(13) & 53396.0 & 17 & 53050-53742 & 6.3 & 162(6)\\ 
1904$-$0150 & 0.379387161967(6) & 0.8898(8) & 53107.0 & 34 & 52752-53461 & 0.5 & 162.2(5)\\ 
\\
1906$+$0414 & 1.043361628655(9) & 11.461(9) & 53677.0 & 8 & 53553-53802 & 0.014 & 349(9)\\ 
1907$+$1149 & 1.42016034143(17) & 159.79(3) & 53445.0 & 26 & 53148-53742 & 3.2 & 202.8(15)\\ 
1910$-$0112 & 1.3606029280(8) & 0.18(6) & 53174.0 & 15 & 52861-53487 & 8.6 & 178(26)\\ 
1910$+$1256 & 0.0049835839397055(12) & 0.00000977(7) & 52970.0 & 183 & 52602-53337 & 0.002 & 38.0650(7)\\ 
1911$+$1347 & 0.0046259624652639(15) & 0.00001712(20) & 53094.0 & 59 & 52718-53471 & 0.020 & 30.99(5)\\ 
\\
1913$+$0904 & 0.163245785775(11) & 17.6167(8) & 53249.0 & 23 & 53004-53494 & 0.1 & 95.3(6)\\ 
1914$+$0219 & 0.457526573507(11) & 1.0181(4) & 53040.0 & 66 & 52608-53472 & 0.6 & 233.8(4)\\ 
1915$+$0227 & 0.3173062332319(16) & 0.29898(15) & 53036.0 & 56 & 52600-53472 & 0.4 & 192.6(5)\\ 
1915$+$1410 & 0.297494121670(7) & 0.0489(8) & 53399.0 & 20 & 53055-53742 & 0.5 & 273.7(3)\\ 
1921$+$0812 & 0.2106484121028(14) & 5.3633(6) & 53277.0 & 39 & 53083-53471 & 0.1 & 84.0(6)\\ 
\\
1927$+$0911 & 0.290305256197(5) & 0.0635(4) & 53454.0 & 44 & 53153-53755 & 0.4 & 202.7(4)\\ 
1941$+$1341 & 0.559084099229(13) & 1.2387(13) & 53041.0 & 59 & 52600-53481 & 1.5 & 147.9(3)\\ 

\hline 
\end{tabular}\end{footnotesize}\end{center}\end{table*}

\begin{table*} 
\caption{Derived parameters for 142 pulsars discovered in this phase
of the Parkes multibeam pulsar survey. Listed are the base-10
logarithms of characteristic age (yr), the surface dipole magnetic
field strength (G), the loss in rotational energy (erg~s$^{-1}$), the
DM-derived distance using the Taylor \& Cordes (1993) model, $D_{\rm
TC}$ (kpc), and the Cordes \& Lazio (2001) model, $D_{\rm CL}$ (kpc),
the corresponding $z$ heights, $z_{TC}$ and $z_{CL}$ (both in kpc),
and the inferred radio luminosity at 1400 MHz for each distance
estimate $L_{TC}$ and $L_{CL}$ (both in mJy~kpc$^2$).}
\label{tb:deriv}
\begin{center}\begin{footnotesize}
\begin{tabular}{rrrrrrrrrr}
\hline 
\multicolumn{1}{c}{PSR J} & 
\multicolumn{1}{c}{$\log[\tau_c]$} & 
\multicolumn{1}{c}{$\log[B]$} & 
\multicolumn{1}{c}{$\log[\dot{E}]$} & 
\multicolumn{1}{c}{$D_{\rm TC}$} & 
\multicolumn{1}{c}{$D_{\rm CL}$} & 
\multicolumn{1}{c}{$z_{\rm TC}$} & 
\multicolumn{1}{c}{$z_{\rm CL}$} & 
\multicolumn{1}{c}{$L_{\rm TC}$} &
\multicolumn{1}{c}{$L_{\rm CL}$} \\ 
\hline 
0733$-$2345 & 6.70 & 12.51 & 31.59 & 12.03 & 8.04 & $-0.43$ & $-0.28$ & 10.1 & 4.5\\ 
0932$-$5327 & 6.92 & 12.79 & 30.59 & 3.91 & 2.31 & $-0.09$ & $-0.06$ & 2.3 & 0.8\\ 
1001$-$5939 & 6.31 & 13.34 & 30.71 & 3.31 & 2.76 & $-0.20$ & $-0.17$ & 1.6 & 1.1\\ 
1020$-$6026 & 5.52 & 11.99 & 34.98 & 30.00 & 12.19 & $-1.49$ & $-0.61$ & 126.0 & 20.8\\ 
1107$-$5907 & 8.65 & 10.68 & 31.34 & 1.81 & 1.28 & $0.04$ & $0.02$ & 0.6 & 0.3\\ 
\\
1125$-$6014 & 10.02 & 8.02 & 33.94 & 1.94 & 1.50 & $0.03$ & $0.02$ & 0.2 & 0.1\\ 
1148$-$5725 & 6.71 & 12.80 & 30.99 & 6.95 & 3.75 & $0.54$ & $0.29$ & 5.8 & 1.7\\ 
1216$-$6410 & 10.54 & 7.88 & 33.15 & 1.71 & 1.33 & $-0.05$ & $-0.04$ & 0.1 & 0.1\\ 
1308$-$5844 & 5.93 & 12.31 & 33.53 & 8.92 & 4.59 & $0.63$ & $0.32$ & 16.7 & 4.4\\ 
1355$-$5747 & 7.66 & 12.09 & 30.52 & 7.57 & 5.08 & $0.53$ & $0.36$ & 20.6 & 9.3\\ 
\\
1357$-$6429 & 3.86 & 12.89 & 36.49 & 4.03 & 2.47 & $-0.18$ & $-0.11$ & 7.1 & 2.7\\ 
1405$-$5641 & 6.91 & 11.94 & 32.30 & 21.25 & 6.73 & $1.76$ & $0.56$ & 45.2 & 4.5\\ 
1439$-$5501 & 9.51 & 9.31 & 32.38 & 0.76 & 0.60 & $0.06$ & $0.05$ & 0.2 & 0.2\\ 
1519$-$5734 & 6.28 & 12.18 & 33.08 & 13.07 & 8.81 & $-0.05$ & $-0.04$ & 76.9 & 34.9\\ 
1519$-$6106 & 6.61 & 12.63 & 31.52 & 7.18 & 4.35 & $-0.40$ & $-0.24$ & 9.8 & 3.6\\ 
\\
1538$-$5732 & 6.08 & 12.10 & 33.65 & 3.88 & 2.83 & $-0.11$ & $-0.08$ & 5.3 & 2.8\\ 
1558$-$5756 & 4.98 & 13.16 & 33.72 & 3.74 & 2.50 & $-0.23$ & $-0.16$ & 2.7 & 1.2\\ 
1600$-$5916 & 7.35 & 12.03 & 31.26 & 6.47 & 3.73 & $-0.54$ & $-0.31$ & 13.8 & 4.6\\ 
1604$-$4718 & 6.23 & 12.21 & 33.11 & 1.48 & 2.14 & $0.10$ & $0.14$ & 0.5 & 1.0\\ 
1611$-$5847 & 9.45 & 10.43 & 30.26 & 2.32 & 1.70 & $-0.22$ & $-0.16$ & 0.6 & 0.3\\ 
\\
1617$-$4608 & 5.74 & 12.49 & 33.56 & 10.98 & 6.07 & $0.60$ & $0.33$ & 18.1 & 5.5\\ 
1622$-$4347 & 6.16 & 12.19 & 33.32 & 5.42 & 3.51 & $0.40$ & $0.26$ & 5.3 & 2.2\\ 
1624$-$4411 & 6.66 & 11.64 & 33.40 & 3.66 & 2.70 & $0.23$ & $0.17$ & 6.4 & 3.5\\ 
1624$-$4613 & $>8.06$ & $<11.51$ & $<30.86$ & 5.18 & 3.76 & $0.20$ & $0.15$ & 10.5 & 5.5\\ 
1627$-$4706 & 6.11 & 11.70 & 34.40 & 7.34 & 6.03 & $0.16$ & $0.13$ & 5.4 & 3.6\\ 
\\
1631$-$4155 & 9.00 & 10.85 & 30.32 & 8.60 & 4.81 & $0.65$ & $0.36$ & 14.1 & 4.4\\ 
1634$-$5107 & 6.71 & 11.96 & 32.68 & 9.20 & 6.09 & $-0.37$ & $-0.24$ & 29.6 & 13.0\\ 
1637$-$4450 & 6.84 & 11.59 & 33.15 & 8.80 & 6.49 & $0.23$ & $0.17$ & 31.0 & 16.8\\ 
1638$-$3815 & 8.16 & 11.37 & 30.95 & 17.25 & 5.66 & $1.73$ & $0.57$ & 184.5 & 19.9\\ 
1638$-$4725 & 6.40 & 12.29 & 32.62 & 6.73 & 6.07 & $-0.04$ & $-0.03$ & 14.5 & 11.8\\ 
\\
1643$-$4505 & 5.07 & 12.44 & 34.97 & 6.37 & 5.64 & $0.06$ & $0.05$ & 11.4 & 8.9\\ 
1646$-$5123 & 6.60 & 12.03 & 32.75 & 10.62 & 5.63 & $-0.73$ & $-0.39$ & 19.2 & 5.4\\ 
1649$-$3805 & 8.05 & 11.00 & 31.90 & 6.89 & 4.30 & $0.50$ & $0.31$ & 47.5 & 18.5\\ 
1650$-$4921 & 6.13 & 11.73 & 34.28 & 5.77 & 4.08 & $-0.31$ & $-0.22$ & 5.3 & 2.7\\ 
1653$-$4030 & 7.57 & 11.83 & 31.20 & 10.94 & 6.68 & $0.40$ & $0.25$ & 47.9 & 17.8\\ 
\\
1654$-$3710 & 7.31 & 11.92 & 31.54 & 14.10 & 6.33 & $0.99$ & $0.44$ & 43.7 & 8.8\\ 
1656$-$3621 & 6.96 & 11.99 & 32.11 & 7.85 & 4.67 & $0.58$ & $0.35$ & 17.9 & 6.3\\ 
1700$-$3611 & 6.74 & 12.41 & 31.71 & 6.80 & 4.48 & $0.44$ & $0.29$ & 23.6 & 10.2\\ 
1700$-$4012 & 7.66 & 11.23 & 32.23 & 5.94 & 5.07 & $0.13$ & $0.11$ & 4.6 & 3.3\\ 
1711$-$4322 & 7.79 & 10.72 & 32.99 & 4.15 & 3.83 & $-0.16$ & $-0.15$ & 4.5 & 3.8\\ 
\\
1715$-$3859 & 6.52 & 12.31 & 32.34 & 10.32 & 9.34 & $-0.06$ & $-0.06$ & 57.5 & 47.1\\ 
1717$-$3953 & 8.72 & 11.28 & 30.00 & 6.68 & 5.65 & $-0.14$ & $-0.11$ & 34.8 & 24.9\\ 
1718$-$4539 & 6.10 & 12.33 & 33.15 & 10.69 & 6.35 & $-0.85$ & $-0.51$ & 9.1 & 3.2\\ 
1722$-$4400 & 6.97 & 11.46 & 33.15 & 7.02 & 5.14 & $-0.53$ & $-0.39$ & 10.8 & 5.8\\ 
1724$-$4500 & 6.77 & 12.34 & 31.79 & 6.15 & 4.56 & $-0.55$ & $-0.41$ & 1.9 & 1.0\\ 
\\
1725$-$2852 & 7.00 & 12.20 & 31.60 & 4.09 & 3.13 & $0.27$ & $0.21$ & 4.2 & 2.4\\ 
1728$-$4028 & 7.77 & 11.66 & 31.15 & 5.40 & 4.08 & $-0.30$ & $-0.23$ & 20.1 & 11.5\\ 
1730$-$2900 & 6.44 & 12.57 & 31.98 & 6.97 & 5.04 & $0.34$ & $0.25$ & 6.3 & 3.3\\ 
1732$-$3426 & 7.12 & 11.57 & 32.63 & 6.35 & 5.74 & $-0.06$ & $-0.05$ & 9.7 & 7.9\\ 
1733$-$2533 & 6.70 & 12.08 & 32.46 & 8.01 & 5.00 & $0.57$ & $0.36$ & 6.4 & 2.5\\ 
\\

\hline 
\end{tabular}\end{footnotesize}\end{center}\end{table*}
\addtocounter{table}{-1}\begin{table*} 
\begin{center}\begin{footnotesize}
\caption{-- {\it continued}}\begin{tabular}{rrrrrrrrrr}
\hline 
\multicolumn{1}{c}{PSR J} & 
\multicolumn{1}{c}{$\log[\tau_c]$} & 
\multicolumn{1}{c}{$\log[B]$} & 
\multicolumn{1}{c}{$\log[\dot{E}]$} & 
\multicolumn{1}{c}{$D_{\rm TC}$} & 
\multicolumn{1}{c}{$D_{\rm CL}$} & 
\multicolumn{1}{c}{$z_{\rm TC}$} & 
\multicolumn{1}{c}{$z_{\rm CL}$} & 
\multicolumn{1}{c}{$L_{\rm TC}$} &
\multicolumn{1}{c}{$L_{\rm CL}$} \\ 
\hline 
1733$-$2837 & 7.04 & 11.97 & 31.98 & 4.72 & 3.87 & $0.20$ & $0.16$ & 1.6 & 1.0\\ 
1734$-$2415 & 6.97 & 11.91 & 32.26 & 3.51 & 2.59 & $0.28$ & $0.21$ & 3.6 & 1.9\\ 
1736$-$2457 & 7.09 & 12.48 & 30.86 & 4.46 & 3.38 & $0.29$ & $0.22$ & 10.3 & 5.9\\ 
1738$-$2330 & 6.56 & 12.62 & 31.64 & 2.79 & 2.05 & $0.21$ & $0.15$ & 3.7 & 2.0\\ 
1739$-$3951 & 8.44 & 10.92 & 31.30 & 1.13 & 0.78 & $-0.09$ & $-0.06$ & 0.2 & 0.1\\ 
\\
1741$-$2719 & 7.83 & 11.23 & 31.88 & 5.99 & 5.28 & $0.17$ & $0.15$ & 7.2 & 5.6\\ 
1742$-$3957 & 8.29 & 11.46 & 30.49 & 6.40 & 3.90 & $-0.57$ & $-0.35$ & 5.7 & 2.1\\ 
1744$-$3922 & 9.25 & 10.22 & 31.08 & 4.60 & 3.06 & $-0.41$ & $-0.27$ & 4.2 & 1.9\\ 
1750$-$2043 & 7.05 & 12.83 & 30.26 & 5.81 & 4.84 & $0.34$ & $0.28$ & 8.1 & 5.6\\ 
1751$-$2857 & 9.74 & 8.33 & 33.87 & 1.44 & 1.10 & $-0.03$ & $-0.02$ & 0.1 & 0.1\\ 
\\
1753$-$1914 & 8.69 & 10.06 & 32.51 & 2.80 & 2.17 & $0.17$ & $0.13$ & 1.0 & 0.6\\ 
1754$-$3510 & 6.90 & 11.75 & 32.71 & 2.22 & 1.95 & $-0.19$ & $-0.17$ & 2.3 & 1.8\\ 
1755$-$1650 & 7.23 & 11.86 & 31.84 & 4.31 & 3.40 & $0.32$ & $0.25$ & 2.4 & 1.5\\ 
1756$-$2251 & 8.65 & 9.74 & 33.23 & 2.92 & 2.48 & $0.05$ & $0.04$ & 5.1 & 3.7\\ 
1758$-$2846 & 8.11 & 11.43 & 30.91 & 1.71 & 1.76 & $-0.07$ & $-0.07$ & 0.6 & 0.6\\ 
\\
1759$-$1736 & 7.70 & 11.66 & 31.30 & 4.83 & 4.00 & $0.26$ & $0.21$ & 2.8 & 1.9\\ 
1801$-$1417 & 10.03 & 8.15 & 33.64 & 1.80 & 1.52 & $0.13$ & $0.11$ & 0.6 & 0.4\\ 
1801$-$3458 & 7.74 & 11.88 & 30.77 & 4.88 & 3.49 & $-0.51$ & $-0.37$ & 2.6 & 1.3\\ 
1802$-$2124 & 9.44 & 8.98 & 33.15 & 3.33 & 2.94 & $0.04$ & $0.03$ & 8.5 & 6.7\\ 
1805$-$1504 & 7.84 & 11.76 & 30.82 & 5.33 & 4.38 & $0.29$ & $0.24$ & 62.5 & 42.2\\ 
\\
1808$-$1020 & 7.09 & 11.84 & 32.15 & 8.95 & 5.29 & $0.72$ & $0.42$ & 18.4 & 6.4\\ 
1808$-$1726 & 8.52 & 10.73 & 31.52 & 9.45 & 7.85 & $0.20$ & $0.16$ & 34.8 & 24.0\\ 
1808$-$2701 & 5.77 & 13.11 & 32.23 & 2.43 & 1.71 & $-0.14$ & $-0.10$ & 0.9 & 0.4\\ 
1811$-$2439 & 7.35 & 11.55 & 32.20 & 3.79 & 3.62 & $-0.20$ & $-0.19$ & 3.7 & 3.4\\ 
1812$-$2526 & 7.45 & 11.38 & 32.34 & 13.76 & 8.00 & $-0.83$ & $-0.48$ & 34.1 & 11.5\\ 
\\
1814$-$0618 & 7.87 & 11.81 & 30.64 & 5.86 & 4.22 & $0.53$ & $0.38$ & 19.9 & 10.3\\ 
1816$-$0755 & 5.73 & 12.08 & 34.40 & 3.24 & 2.78 & $0.23$ & $0.20$ & 1.8 & 1.3\\ 
1817$-$0743 & $>8.92$ & $<10.79$ & $<30.59$ & 0.77 & 0.80 & $0.05$ & $0.05$ & 0.1 & 0.2\\
1819$-$1318 & 7.60 & 11.98 & 30.83 & 1.52 & 1.15 & $0.02$ & $0.02$ & 0.4 & 0.2\\ 
1820$-$0509 & 6.76 & 11.75 & 32.98 & 2.84 & 2.51 & $0.22$ & $0.20$ & 2.0 & 1.6\\ 
\\
1821$-$0256 & 8.25 & 11.10 & 31.32 & 2.55 & 2.36 & $0.24$ & $0.22$ & 1.2 & 1.1\\ 
1822$-$0848 & 7.99 & 11.53 & 30.96 & 4.24 & 3.77 & $0.17$ & $0.15$ & 0.7 & 0.6\\ 
1824$-$0127 & 7.00 & 12.50 & 31.00 & 2.21 & 1.88 & $0.20$ & $0.17$ & 2.9 & 2.1\\ 
1824$-$2233 & 7.79 & 11.78 & 30.88 & 4.22 & 3.78 & $-0.33$ & $-0.29$ & 3.9 & 3.1\\ 
1824$-$2328 & 7.14 & 12.21 & 31.30 & 5.61 & 4.55 & $-0.48$ & $-0.39$ & 10.1 & 6.6\\ 
\\
1827$-$0750 & 6.44 & 11.82 & 33.49 & 7.45 & 6.04 & $0.23$ & $0.19$ & 77.7 & 51.1\\ 
1828$-$2119 & 6.81 & 11.91 & 32.57 & 12.06 & 6.77 & $-0.99$ & $-0.56$ & 55.3 & 17.4\\ 
1829$+$0000 & 6.78 & 11.51 & 33.41 & 3.40 & 3.22 & $0.29$ & $0.27$ & 5.0 & 4.5\\ 
1830$-$0052 & 7.36 & 11.46 & 32.36 & 7.37 & 5.68 & $0.55$ & $0.42$ & 2.2 & 1.3\\ 
1830$-$0131 & 6.06 & 11.76 & 34.36 & 2.69 & 2.66 & $0.19$ & $0.19$ & 2.5 & 2.5\\ 
\\
1830$-$1414 & 8.20 & 11.39 & 30.81 & 7.80 & 6.26 & $-0.25$ & $-0.20$ & 6.1 & 3.9\\ 
1831$-$0952 & 5.11 & 11.88 & 36.04 & 4.32 & 4.04 & $-0.01$ & $-0.01$ & 6.2 & 5.4\\ 
1832$+$0029 & 6.75 & 11.96 & 32.59 & 1.45 & 1.32 & $0.11$ & $0.10$ & 0.3 & 0.2\\ 
1834$-$0031 & 7.06 & 11.59 & 32.69 & 4.47 & 3.96 & $0.27$ & $0.24$ & 3.4 & 2.7\\ 
1834$-$1855 & 7.08 & 12.23 & 31.38 & 5.95 & 4.62 & $-0.52$ & $-0.40$ & 17.0 & 10.2\\ 
\\
1835$-$0349 & 6.64 & 12.21 & 32.30 & 5.27 & 5.59 & $0.17$ & $0.18$ & 4.4 & 5.0\\ 
1835$-$0944 & 5.72 & 11.91 & 34.75 & 4.63 & 4.37 & $-0.08$ & $-0.08$ & 8.8 & 7.8\\ 
1835$-$0946 & 8.14 & 11.11 & 31.49 & 4.02 & 3.61 & $-0.07$ & $-0.06$ & 2.9 & 2.3\\ 
1835$-$1548 & 6.79 & 12.04 & 32.36 & 13.52 & 7.51 & $-0.89$ & $-0.50$ & 11.0 & 3.4\\ 
1836$-$1324 & 6.44 & 11.64 & 33.86 & 3.87 & 3.48 & $-0.20$ & $-0.18$ & 1.5 & 1.2\\ 
\\

\hline 
\end{tabular}\end{footnotesize}\end{center}\end{table*}
\addtocounter{table}{-1}\begin{table*} 
\begin{center}\begin{footnotesize}
\caption{-- {\it continued}}\begin{tabular}{rrrrrrrrrr}
\hline 
\multicolumn{1}{c}{PSR J} & 
\multicolumn{1}{c}{$\log[\tau_c]$} & 
\multicolumn{1}{c}{$\log[B]$} & 
\multicolumn{1}{c}{$\log[\dot{E}]$} & 
\multicolumn{1}{c}{$D_{\rm TC}$} & 
\multicolumn{1}{c}{$D_{\rm CL}$} & 
\multicolumn{1}{c}{$z_{\rm TC}$} & 
\multicolumn{1}{c}{$z_{\rm CL}$} & 
\multicolumn{1}{c}{$L_{\rm TC}$} &
\multicolumn{1}{c}{$L_{\rm CL}$} \\ 
\hline 
1837$+$0053 & 8.30 & 11.13 & 31.15 & 3.27 & 3.41 & $0.20$ & $0.21$ & 3.6 & 4.0\\ 
1839$-$0436 & 6.47 & 11.55 & 33.98 & 4.98 & 5.45 & $0.06$ & $0.06$ & 5.7 & 6.8\\ 
1839$-$1238 & 6.79 & 12.49 & 31.45 & 4.13 & 3.74 & $-0.23$ & $-0.21$ & 6.3 & 5.2\\ 
1840$+$0214 & 6.18 & 12.41 & 32.81 & 5.50 & 4.93 & $0.33$ & $0.30$ & 2.1 & 1.7\\ 
1840$-$0840 & 6.55 & 13.05 & 30.79 & 4.84 & 4.82 & $-0.14$ & $-0.14$ & 23.4 & 23.2\\ 
\\
1840$-$1207 & 6.57 & 12.20 & 32.46 & 8.82 & 6.13 & $-0.49$ & $-0.34$ & 17.1 & 8.3\\ 
1841$+$0130 & 7.76 & 10.20 & 34.08 & 3.20 & 3.59 & $0.16$ & $0.18$ & 0.6 & 0.8\\ 
1841$-$1404 & 7.52 & 11.97 & 31.04 & 10.42 & 6.36 & $-0.77$ & $-0.47$ & 19.5 & 7.3\\ 
1842$+$0257 & 6.22 & 12.99 & 31.60 & 4.00 & 4.26 & $0.23$ & $0.25$ & 4.2 & 4.7\\ 
1842$+$0358 & 6.66 & 11.64 & 33.40 & 3.21 & 3.53 & $0.22$ & $0.24$ & 0.9 & 1.1\\ 
\\
1842$+$0638 & 7.81 & 11.19 & 31.99 & 11.23 & 6.80 & $0.99$ & $0.60$ & 26.5 & 9.7\\ 
1843$-$1507 & 6.11 & 12.32 & 33.15 & 8.71 & 5.52 & $-0.78$ & $-0.49$ & 12.9 & 5.2\\ 
1845$+$0623 & 7.62 & 11.95 & 30.88 & 3.53 & 3.86 & $0.27$ & $0.29$ & 4.1 & 4.9\\ 
1845$-$0826 & 6.03 & 12.39 & 33.15 & 4.86 & 4.48 & $-0.21$ & $-0.19$ & 7.8 & 6.6\\ 
1845$-$1114 & 6.21 & 11.81 & 33.95 & 5.54 & 4.52 & $-0.37$ & $-0.30$ & 16.0 & 10.6\\ 
\\
1845$-$1351 & 6.63 & 12.71 & 31.32 & 7.01 & 4.95 & $-0.60$ & $-0.42$ & 16.2 & 8.1\\ 
1846$-$0749 & 6.64 & 11.83 & 33.08 & 9.82 & 7.16 & $-0.42$ & $-0.30$ & 33.8 & 17.9\\ 
1848$-$0601 & 7.09 & 11.41 & 33.00 & 11.55 & 9.52 & $-0.41$ & $-0.34$ & 32.0 & 21.8\\ 
1848$-$1150 & 7.16 & 12.14 & 31.40 & 4.73 & 3.98 & $-0.38$ & $-0.32$ & 4.7 & 3.3\\ 
1849$+$0409 & 5.75 & 12.61 & 33.28 & 2.41 & 2.42 & $0.10$ & $0.10$ & 0.6 & 0.6\\ 
\\
1851$-$0114 & 6.78 & 12.19 & 32.04 & 6.26 & 6.90 & $-0.06$ & $-0.06$ & 11.0 & 13.3\\ 
1853$+$0853 & 7.08 & 12.66 & 30.53 & 7.67 & 6.75 & $0.48$ & $0.43$ & 6.5 & 5.0\\ 
1853$+$1303 & 9.87 & 8.29 & 33.71 & 1.60 & 2.05 & $0.15$ & $0.19$ & 1.0 & 1.7\\ 
1856$-$0526 & 6.54 & 11.90 & 33.11 & 3.41 & 3.32 & $-0.21$ & $-0.21$ & 4.7 & 4.4\\ 
1901$-$0315 & 6.69 & 12.16 & 32.30 & 6.84 & 5.94 & $-0.44$ & $-0.38$ & 4.2 & 3.2\\ 
\\
1901$+$0435 & 6.10 & 12.39 & 33.00 & 30.00 & 19.07 & $-0.08$ & $-0.05$ & 243.0 & 98.2\\ 
1901$+$0621 & 8.86 & 11.09 & 30.08 & 3.11 & 3.40 & $0.04$ & $0.05$ & 4.5 & 5.4\\ 
1903$-$0258 & 6.85 & 11.66 & 32.99 & 3.07 & 3.16 & $-0.22$ & $-0.22$ & 1.3 & 1.4\\ 
1903$+$0925 & 5.18 & 12.56 & 34.51 & 4.20 & 4.95 & $0.11$ & $0.13$ & 3.5 & 4.9\\ 
1904$-$0150 & 6.83 & 11.77 & 32.81 & 4.77 & 4.36 & $-0.32$ & $-0.29$ & 2.0 & 1.7\\ 
\\
1906$+$0414 & 6.16 & 12.54 & 32.60 & 9.24 & 7.50 & $-0.24$ & $-0.20$ & 19.6 & 12.9\\ 
1907$+$1149 & 5.15 & 13.18 & 33.34 & 5.33 & 5.98 & $0.17$ & $0.19$ & 3.7 & 4.6\\ 
1910$-$0112 & 8.08 & 11.70 & 30.45 & 6.36 & 5.03 & $-0.53$ & $-0.42$ & 2.8 & 1.8\\ 
1910$+$1256 & 9.91 & 8.35 & 33.49 & 1.95 & 2.32 & $0.06$ & $0.07$ & 1.9 & 2.7\\ 
1911$+$1347 & 9.63 & 8.45 & 33.83 & 1.60 & 2.07 & $0.05$ & $0.07$ & 0.2 & 0.3\\ 
\\
1913$+$0904 & 5.17 & 12.24 & 35.20 & 3.48 & 2.85 & $-0.04$ & $-0.03$ & 0.8 & 0.6\\ 
1914$+$0219 & 6.85 & 11.84 & 32.62 & 9.57 & 6.98 & $-0.67$ & $-0.49$ & 68.7 & 36.5\\ 
1915$+$0227 & 7.23 & 11.49 & 32.57 & 7.00 & 5.89 & $-0.50$ & $-0.42$ & 27.0 & 19.1\\ 
1915$+$1410 & 7.98 & 11.09 & 31.86 & 7.81 & 7.93 & $0.16$ & $0.17$ & 6.1 & 6.3\\ 
1921$+$0812 & 5.79 & 12.03 & 34.36 & 3.44 & 3.46 & $-0.18$ & $-0.18$ & 7.8 & 7.9\\ 
\\
1927$+$0911 & 7.86 & 11.14 & 32.00 & 7.05 & 6.75 & $-0.45$ & $-0.44$ & 8.0 & 7.3\\ 
1941$+$1341 & 6.85 & 11.93 & 32.45 & 8.52 & 5.50 & $-0.66$ & $-0.43$ & 12.3 & 5.1\\ 

\hline 
\end{tabular}\end{footnotesize}\end{center}\end{table*}

\clearpage
\begin{table*}
\caption{Summary of the 13 binary pulsars discovered in this phase of
the Parkes multibeam survey. For each system, we give the spin period
($P$), orbital period ($P_b$), projected semi-major axis ($x$) and
orbital eccentricity ($e$).  The right-hand column gives the source of
the more detailed parameters which either refer to external papers or
additional tables below.}
\begin{tabular}{lllllr} \hline
PSR            & $P$    &   $P_b$    &    $x$   & $e$ & Further details\\
               &  (ms)  & (days)  & (lt sec)&     &     \\\hline
J1125$-$6014     & 2.63 &    8.75   &    8.34 & $7.9 \times 10^{-7}$  & Table \ref{tb:elbin}  \\
J1216$-$6410     & 3.54 &    4.04   &    2.94 & $6.8 \times 10^{-6}$  & Table \ref{tb:elbin}  \\
J1439$-$5501     & 28.6 &    2.12   &    9.84 & $5.0 \times 10^{-5}$  & Table \ref{tb:elbin} \\
J1638$-$4725     & 764  & 1940      & 2380    & 0.95                  & Lyne et al.~(in preparation)\\
J1711$-$4322     & 103  &  922      &  140    & $2.4 \times 10^{-3}$  & Table \ref{tb:btbin} \\
J1744$-$3922     & 172  &    0.191  &    0.212& $1.3 \times 10^{-3}$  & \citet{fsk+04}\\
J1751$-$2857     & 3.92 &  111      &   32.5  & $1.3 \times 10^{-4}$  & \citet{sfl+05}\\
J1756$-$2251     & 28.5 &    0.320  &    2.76 & 0.18                  & \citet{fkl+05}\\
J1802$-$2124     & 12.6 &    0.699  &    3.72 & $3.2 \times 10^{-6}$  & \citet{fsk+04}\\
J1822$-$0848     & 835  &  287      &   97.8  & 0.059                 & Table \ref{tb:btbin}\\
J1841$+$0130     & 29.8 &   10.5    &    3.50 & $8.2 \times 10^{-5}$  & Table \ref{tb:elbin}\\
J1853$+$1303     & 4.09 &  116      &   40.8  & $2.4 \times 10^{-5}$  & \citet{sfl+05}\\
J1910$+$1256     & 4.98 &   58.5    &   21.1  & $2.3 \times 10^{-4}$  & \citet{sfl+05}\\\hline
\end{tabular}
\label{tb:allbin}
\end{table*}

\begin{table*}
\caption{Orbital parameters for the two binary pulsars obtained using
the Blandford \& Teukolsky (1976) binary model.  The minimum companion
mass is calculated by assuming an inclination angle of $90^\circ$ and
a neutron star mass of 1.35\,M$_\odot$. Figures in parentheses
represent 1-$\sigma$ uncertainties in the least significant digit(s).}
\begin{tabular}{lll} \hline
PSR                                         &J1711$-$4322&J1822$-$0848 \\\hline
Orbital period (d)                          & 922.4707(7)& 286.8303(14)\\
Projected semi-major axis of orbit (lt sec) & 139.6245(6)& 97.7279(20) \\
Eccentricity                                & 0.002375(6)& 0.058962(9) \\
Epoch of periastron (MJD)                   & 50208.8(3) & 52672.848(9)\\
Longitude of periastron (degrees)           & 293.75(12) & 272.583(10) \\
Mass function (M$_\odot$)                   & 0.00343446(2)&0.0121811(3)\\
Minimum companion mass (M$_\odot$)          & 0.20       & 0.32  \\ \hline
\end{tabular} 
\label{tb:btbin}
\end{table*}

\begin{table*}
\caption{Orbital parameters for the four binary pulsars obtained using
the `ELL1' binary timing model \citep{lcw+01}, where the first and
second Laplace-Lagrange parameters are defined by $e \cos \omega$ and
$e \sin \omega$ respectively (for an orbital eccentricity $e$ and
longitude of periastron $\omega$).  The minimum companion mass is
calculated by assuming an inclination angle of $90^\circ$ and a
neutron star mass of 1.35\,M$_\odot$.  Figures in parentheses
represent 1-$\sigma$ uncertainties in the least significant digit(s).}
\begin{tabular}{lllll} \hline
PSR                                         & J1125$-$6014   & J1216$-$6410 &
J1439$-$5501 & J1841$+$0130\\\hline
Orbital period (d)                          & 8.75260353(5)&4.03672718(6)&
2.117942520(3)&10.471626(5) \\
Projected semi-major axis of orbit (lt sec) & 8.339198(5)  &2.937091(7)  &
9.833404(7)&3.50409(18)\\
First Laplace-Lagrange parameter            &  $-$0.0000008(13) &0.0000004(60)   &
$-$0.0000495(15)&$-$0.0000033(10)\\
Second Laplace-Lagrange parameter           & 0.00000005(120) & $-$0.000007(6)  &
0.0000059(13)&$-$0.00008(10)\\
Epoch of ascending node (MJD)     & 53171.5856408(11)&53055.3611070(20)&
53058.0499527(3)&52747.55445(11)\\
Mass function (M$_\odot$)                   & 0.008127952(7)&0.001669461(6)&0.2275967(2)&0.00042129(3)\\
Minimum companion mass (M$_\odot$)   & 0.28         & 0.16& 1.11& 0.096
\\ \hline
\end{tabular} 
\label{tb:elbin}
\end{table*}

\begin{table*}
\caption{Summary of the analytic function fits to models S and C.  The
various parameters used are described in
Equations~\ref{eq:rdist}--\ref{eq:pdist}. Figures in parentheses give
the 1-$\sigma$ uncertainty in the least significant digit.}
\begin{tabular}{crrl}\hline
Parameter & \multicolumn{1}{c}{Model S fit}   & 
\multicolumn{1}{c}{Model C fit} & \multicolumn{1}{c}{Unit} \\
\hline
  A       & 44(7)         &  41(5)      & kpc$^{-2}$\\
  B       & 0.2(2)        &  1.9(3)     &      \\
  C       & 1.4(6)        &  5.0(6)       &      \\
  D       & 0.39(2)       &  0.75(3)    &      \\
  E       & 0.33(3)       &  0.18(1)    & kpc  \\
  F       & --0.59(5)     &  --0.77(7)  &      \\
  G       & 3.45(8)       &  3.5(1)     &      \\
  H       & 0.51(1)       &  0.52(1)    &      \\
  I       & 2.71(1)       &  2.70(1)    &      \\
  J       & --0.34(1)     &--0.34(1)    &      \\
\hline
\end{tabular}
\label{tb:fits}
\end{table*}

\begin{table*}
\caption{A comparison of the birth rates derived from the pulsar
current analysis as a function of luminosity cut-off and beaming model.
Figures in parentheses represent statistical 1-$\sigma$ uncertainties
in the least significant digit(s) calculated
as the square root of the sum of the squares of the
scale factors in the maximum pulsar current bin \citep{vn81}.}
\begin{tabular}{rrrr}\hline
\multicolumn{1}{c}{Beaming model}
& \multicolumn{3}{c}{Minimum luminosity (mJy kpc$^2$)}\\
              &
\multicolumn{1}{c}{0.1} &
\multicolumn{1}{c}{1} &
\multicolumn{1}{c}{10} \\
\hline
No correction (i.e. $f=1$)&  0.34(5) & 0.28(4) & 0.12(3) \\
\citet{tm98} & 1.38(21) & 1.14(18) & 0.42(3) \\
\citet{big90b} & 1.14(18) & 0.94(15) & 0.35(8) \\
\citet{lm88} & 1.32(21) & 1.09(17) & 0.41(9) \\
\citet{nv83} & 0.51(8) & 0.43(7) & 0.16(4) \\
\hline
\end{tabular}
\label{tb:br}
\end{table*}

\begin{table*}
\caption{A comparison of the number of detected pulsars for the real
sample with several models developed in this paper for the various
pulsar surveys.  The completed pulsar surveys are the Parkes multibeam
survey (this paper, PMB), the Swinburne intermediate latitude survey
\citep[][SIL]{ebvb01}, the Parkes high latitude survey
\citep[][PH]{bjd+06}, the Swinburne high latitude survey
\citep[][SHL]{jac05}. Pulsar surveys which are currently on-going are
the Perseus arm (PA) survey with the Parkes multibeam system and the
PALFA surveys with Arecibo \citep{cfl+06}. PALFAi refers to the inner
Galaxy survey, while PALFAa refers to the anticentre survey. For the
completed pulsar surveys, the column labeled `Real' lists the number
of pulsars detected (i.e.~discoveries and redetections). For the
on-going surveys the numbers marked with a * refer to currently known
pulsars in the survey regions with flux densities higher than the
nominal survey threshold. The remaining columns are the average number
of predicted detections using each model.}
\begin{tabular}{rrrrr}\hline
Survey & Real & Model S & Model C & Model C$^{\prime}$ \\
\hline
\multicolumn{5}{c}{Completed pulsar surveys}\\ 
PMB    & 986  &  967    &  991    & 984 \\
SIL    & 156  &  146    &   88    & 160 \\
PH     &  32  &   33    &   23    &  30 \\
SHL    &  51  &   52    &   17    &  38 \\
\hline
\multicolumn{5}{c}{On-going pulsar surveys}\\ 
PA     &  17* &   62    &  35     &  32 \\
PALFAi & 200* &  480    &  550    &  530\\
PALFAa &   6* &   54    &  30     &   30\\
\hline
\end{tabular}
\label{tb:yields}
\end{table*}

\end{document}